\documentclass[journal]{IEEEtran}
\usepackage[intlimits]{amsmath}
\usepackage{lettrine}
\usepackage{stfloats}
\usepackage{graphicx}
\usepackage{subfigure}
\usepackage{textcomp}
\usepackage{amssymb,amsmath}
\usepackage{algorithm,algorithmic}
\usepackage[english]{babel}
\usepackage{lscape,graphicx,amsfonts}
\usepackage[intlimits]{amsmath}
\usepackage{import}
\usepackage{textcomp}						
\usepackage{algorithm,algorithmic}
\usepackage{longtable}
\usepackage[Lenny]{fncychap}
\usepackage{fancyhdr}
\usepackage{subfigure}
\usepackage{wrapfig}
\usepackage{graphicx, subfigure}
\usepackage{multirow}
\usepackage{color,soul}
\usepackage[table]{xcolor}
\newcolumntype{s}{>{\columncolor[HTML]{FE6F5E}} c}
\newcolumntype{t}{>{\columncolor[HTML]{5D8AA8}} c}

\definecolor{darkblue}{rgb}{0,0.39,0.53}
\definecolor{darkred}{rgb}{0.76,0.26,0.31}
\usepackage[colorinlistoftodos]{todonotes}

\begin{document}

\title{\begin{centering}{\fontsize{35}{35}\fontfamily{bch}\selectfont\textbf{Deep learning in Ultrasound Imaging}}\end{centering}}

\author{\begin{centering}
\vspace{-5pt}	
{\fontsize{13}{14}\fontfamily{bch}\selectfont  \textit{Deep learning is taking an ever more prominent role in medical imaging. This paper discusses applications of this powerful approach in ultrasound imaging systems along with domain-specific opportunities and challenges.}} \\
\vspace{10pt}
{
\thanks{\fontsize{9}{9}\fontfamily{bch}\selectfont \textbf{R. J. G. van Sloun} is with the department of Electrical Engineering, Eindhoven University of Technology, Eindhoven, The Netherlands (email: r.j.g.v.sloun@tue.nl)}
\fontsize{11}{13}\fontfamily{bch}\selectfont \sc Ruud J.G. van Sloun$^{1}$,  Regev Cohen$^{2}$ and Yonina C. Eldar$^{3}$}
\end{centering}
\thanks{\fontsize{9}{9}\fontfamily{bch}\selectfont \textbf{R. Cohen} is with the department of Electrical Engineering, Technion, Israel}
\thanks{\fontsize{9}{9}\fontfamily{bch}\selectfont \textbf{Y. C. Eldar} is with the Faculty of Mathematics and Computer science, Weizmann Institute of Science, Rehovot, Israel (email: yonina.eldar@weizmann.ac.il)}
\thanks{}
\thanks{\fontsize{8}{8}\fontfamily{bch}\selectfont Accepted for publication in the \textit{Proceedings of the IEEE}. \textcopyright 2019 IEEE.  Personal use of this material is permitted.  Permission from IEEE must be obtained for all other uses, in any current or future media, including reprinting/republishing this material for advertising or promotional purposes, creating new collective works, for resale or redistribution to servers or lists, or reuse of any copyrighted component of this work in other works.”}
}

\maketitle
\maketitle
%\pagestyle{empty} \thispagestyle{empty}
%\onecolumn
%\begin{abstract}
\noindent{\boldmath
\textbf{
\fontsize{9}{9}\fontfamily{bch}\selectfont
{\color{darkred} ABSTRACT} $\vert$ We consider deep learning strategies in ultrasound systems, from the front-end to advanced applications. Our goal is to provide the reader with a broad understanding of the possible impact of deep learning methodologies on many aspects of ultrasound imaging. In particular, we discuss methods that lie at the interface of signal acquisition and machine learning, exploiting both data structure (e.g. sparsity in some domain) and data dimensionality (big data) already at the raw radio-frequency channel stage. As some examples, we outline efficient and effective deep learning solutions for adaptive beamforming and adaptive spectral Doppler through artificial agents, learn compressive encodings for color Doppler, and provide a framework for structured signal recovery by learning fast approximations of iterative minimization problems, with applications to clutter suppression and super-resolution ultrasound. These emerging technologies may have considerable impact on ultrasound imaging, showing promise across key components in the receive processing chain.}}
%\end{abstract}

\vspace{10pt}
\noindent{\boldmath
\textbf{
\fontsize{9}{9}\fontfamily{bch}\selectfont
\noindent {\color{darkred} KEYWORDS} $\vert$ Deep learning; ultrasound imaging; image reconstruction; beamforming, Doppler, compression, deep unfolding, super resolution.}} 

\section{Introduction}
\label{sec:Introduction}
\noindent Diagnostic imaging plays a critical role in healthcare, serving as a fundamental asset for timely diagnosis, disease staging and management as well as for treatment choice, planning, guidance, and follow-up. Among the diagnostic imaging options, ultrasound imaging \cite{szabo2004diagnostic} is uniquely positioned, being a highly cost-effective modality that offers the clinician an unmatched and invaluable level of interaction, enabled by its real-time nature. Its portability and cost-effectiveness permits point-of-care imaging at the bedside, in emergency settings, rural clinics, and developing countries. Ultrasonography is increasingly used across many medical specialties, spanning from obstetrics to cardiology and oncology, and its market share is globally growing. 

On the technological side, ultrasound probes are becoming increasingly compact and portable, with the market demand for low-cost `pocket-sized' devices expanding \cite{baran2009design,chernyakova2014fourier}. Transducers are miniaturized, allowing e.g. in-body imaging for interventional applications. At the same time, there is a strong trend towards 3D imaging \cite{provost20143d} and the use of high-frame-rate imaging schemes \cite{tanter2014ultrafast}; both accompanied by dramatically increasing data rates that pose a heavy burden on the probe-system communication and subsequent image reconstruction algorithms. Systems today offer a wealth of advanced applications and methods, including shear wave elasticity imaging \cite{bercoff2004supersonic}, ultra-sensitive Doppler \cite{demene2015spatiotemporal}, and ultrasound localization microscopy for super-resolution microvascular imaging \cite{errico2015ultrafast}.

With the demand for high-quality image reconstruction and signal extraction from unfocused planar wave transmissions that facilitate fast imaging, and a push towards miniaturization, modern ultrasound imaging leans heavily on innovations in powerful receive channel processing.  In this paper, we discuss how artificial intelligence and deep learning methods can play a compelling role in this process, and demonstrate how these data-driven systems can be leveraged across the ultrasound imaging chain. We aim to provide the reader with a broad understanding of the possible impact of deep learning on a variety of ultrasound imaging aspects, placing particular emphasis on methods that exploit both the power of data and signal structure (for instance sparsity in some domain) to yield robust and data-efficient solutions. We believe that methods that exploit models and structure together with learning from data can pave the way to interpretable and powerful processing methods from limited training sets. As such, throughout the paper, we will typically first discuss an appropriate model-based solution for the problems considered, and then follow by a data-driven deep learning solution derived from it.

We start by briefly describing a standard ultrasound imaging chain in Section~\ref{sec:signalmodel}. We then elaborate on several dedicated deep learning solutions that aim at improving key components in this processing pipeline, covering adaptive beamforming (Section~\ref{sec:deep_beamforming}), adaptive spectral Doppler (Section~\ref{sec:spectraldoppler}), compressive tissue Doppler (Section~\ref{sec:DopplerNet}), and clutter suppression (Section~\ref{sec:RPCA}). In Section~\ref{sec:deepulm}, we show how the synergetic exploitation of deep learning and signal structure enables robust super-resolution microvascular ultrasound imaging. Finally, we discuss future perspectives, opportunities, and challenges for the holistic integration of artificial intelligence and deep learning methods in ultrasound systems.

\section{The Ultrasound Imaging Chain \\ at a glance}
\label{sec:signalmodel}
\subsection{Transmit schemes}
\noindent The resolution, contrast, and overall fidelity of ultrasound pulse-echo imaging relies on careful optimization across its entire imaging chain. At the front-end, imaging starts with the design of appropriate transmit schemes. 

At this stage, crucial trade-offs are made, in which the frame rate, imaging depth, and attainable axial and lateral resolution are weighted carefully against each other: improved resolution can be achieved through the use of higher pulse modulation frequencies; yet, these shorter wavelengths suffer from increased absorption and thus lead to reduced penetration depth. Likewise, high frame rate can be reached by exploiting parallel transmission schemes based on e.g. planar or diverging waves. However, use of such unfocused transmissions comes at the cost of loss in lateral resolution compared to line-based scanning with tightly focused beams. As such, optimal transmit schemes depend on the application. 

Today, an increasing amount of ultrasound applications rely on high frame-rate (dubbed \textit{ultrafast}) imaging. Among these are e.g. ultrasound localization microscopy (see Section~\ref{sec:deepulm}), highly-sensitive Doppler, and shear wave elastography. Where the former two mostly exploit the incredible vastness of data to obtain accurate signal statistics, the later leverages high-speed imaging to track ultrasound-induced shear waves propagating at several meters per second. 

With the expanding use of ultrafast transmit sequences in modern ultrasound imaging, a strong burden is placed on the subsequent receive channel processing. High data-rates not only raise substantial hardware complications related to power consumption, data storage and data transfer, the corresponding unfocused transmissions require much more advanced receive beamforming and clutter suppression to reach satisfactory image quality. 

%- beampattern design (apodization, delays, focused, plane-wave, diverging wave)
%- transmission sequences (e.g. ensembles for Doppler)

\subsection{Receive processing, sampling and beamforming}
\noindent Modern receive channel processing is shifting towards the digital domain, relying on computational power and very-high-bandwidth communication channels to enable advanced digital parallel (pixel-based) beamforming and coherent compounding across multiple transmit/receive events. For large channel counts, e.g. in dense matrix probes that facilitate high-resolution 3D imaging, the number of coaxial cables required to connect all probe elements to the back-end system quickly becomes infeasible. To address this, dedicated switching and processing already takes place in the probe head, e.g. in the form of multiplexing or microbeamforming. Slow-time\footnote{In ultrasound imaging we make a distinction between slow- and fast-time: slow-time refers to a sequence of snapshots (i.e., across multiple transmit/receive events), at the pulse repetition rate, whereas fast-time refers to samples along depth.} multiplexing distributes the received channel data across multiple transmits, by only communicating a subset of the number of channels to the back-end for each such transmit. This consequently reduces the achieved frame rate. In microbeamforming, an analog pre-beamforming step is performed to compress channel data from multiple (adjacent) elements into a single focused line. This however impairs flexibility in subsequent digital beamforming, limiting the achievable image quality. Other approaches aim at mixing multiple channels through analog modulation with chipping sequences \cite{besson2018compressive}. Additional analog processing includes signal amplification by a low-noise amplifier (LNA) as well as depth (i.e. fast-time) dependent gain compensation (TGC) for attenuation correction. 

Digital receive beamforming in ultrasound imaging is dynamic, i.e. receive focusing is dynamically optimized based on the scan depth. The industry standard is delay-and-sum beamforming, where depth-dependent channel tapering (or apodization) is optimized and fine-tuned based on the system and application. Delay-and-sum beamforming is commonplace due to its low complexity, providing real-time image reconstruction, albeit at a high sampling rate and non-optimal image quality.  

Performing beamforming in the digital domain requires sampling the signals received at the transducer elements and transmitting the samples to a back-end processing unit. To achieve sufficient delay resolution for focusing, the received signals are typically sampled at 4-10 times their bandwidth, i.e., the sampling rate may severely exceed the Nyquist rate. A possible approach for sampling rate reduction is to consider the received signals within the framework of finite rate of innovation (FRI) \cite{eldar2015sampling,gedalyahu2011multichannel}. Tur \textit{et al.} \cite{tur2011innovation} modeled the received signal at each element as a finite sum of replicas of the transmitted pulse backscattered from reflectors. The replicas are fully described by their unknown amplitudes and delays, which can be recovered from the signals' Fourier series coefficients. The latter can be computed from low-rate samples of the signal using compressed sensing (CS) techniques \cite{eldar2015sampling,eldar2012compressed}. In \cite{wagner2011xampling,wagner2012compressed}, the authors extended this approach and introduce compressed beamforming. It was shown that the beamformed signal follows an FRI model and thus it can be reconstructed from a linear combination of the Fourier coefficients of the received signals. Moreover, these coefficients can be obtained from low-rate samples of the received signals taken according to the Xampling framework \cite{mishali2011xampling2,mishali2011xampling,michaeli2012xampling}. Chernyakova \textit{et al.} showed this Fourier domain relationship between the beam and the received signals holds irrespective of the FRI model. This leads to a general concept of frequency domain beamforming (FDBF) \cite{chernyakova2014fourier} which is equivalent to beamforming in time. FDBF allows to sample the received signals at their effective Nyquist rate without assuming a structured model, thus, it avoids the oversampling dictated by digital implementation of beamforming in time. Furthermore, when assuming that the beam obeys a FRI model, the received signals can be sampled at sub-Nyquist rates, leading to up to 28 fold reduction in sampling rate \cite{chernyakova2018fourier,burshtein2016sub,lahav2017focus}. \\

\begin{figure*}[t!]
	\centering
	\includegraphics[trim=0cm 11cm 10.5cm 0cm, clip=true,width=500px]{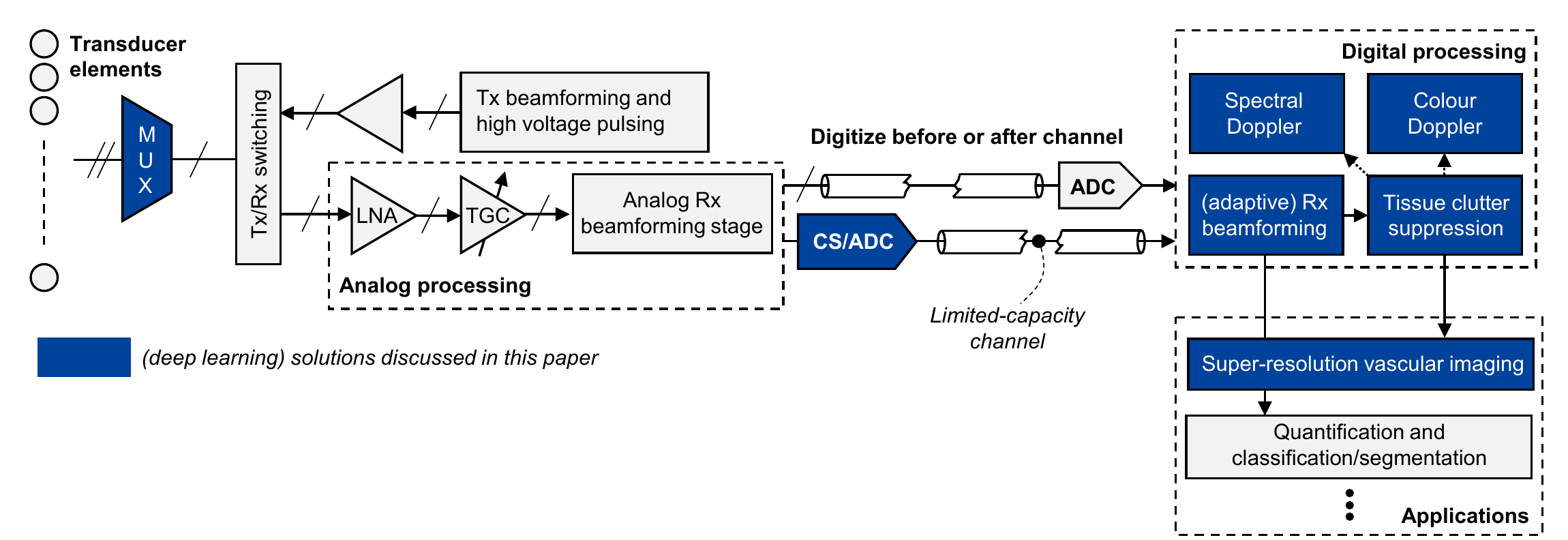}
	\caption{Overview of the ultrasound imaging chain, along with the deep learning solutions discussed in this paper. Note that, today, analog processing at the front-end typically comprises some form of lossy (micro-)beamforming to reduce data rates, in contrast to the here advocated paradigm based on compressive sub-Nyquist sampling, intelligent ASICs with neural edge computing, and subsequent remote deep-learning-based processing of low-rate channel data.}
	\label{fig:overview}
\end{figure*}

\subsection{B-mode, M-mode, and Doppler}
\noindent 
Ultrasound imaging provides anatomical information through the so-called Brightness-mode (B-mode). B-mode imaging is performed by envelope-detecting the beamformed signals, e.g. through calculation of the magnitude of the complex in-phase and quadrature (IQ) data. For visualization purposes, the dynamic range of these envelope-detected signals is subsequently compressed via a logarithmic transformation, or specifically-designed compression curves based on a look-up table. Scan conversion then maps these intensities to the desired (Cartesian) pixel coordinate system. The visualization of a single B-mode scan line (i.e. brightness over fast time) across multiple transmit-receive events (i.e. slow-time), is called motion-mode (M-mode) imaging. 

Beyond anatomical information, ultrasound imaging also permits the measurement of functional parameters related to blood flow and tissue displacement. The extraction of such velocity signals is called Doppler processing. We distinguish between two types of velocity estimators: Color Doppler and Spectral Doppler. Color Doppler provides an estimate of the mean velocity through evaluation of the first lag of the autocorrelation function for a series of snapshots across slow-time \cite{loupas1995axial}. Spectral Doppler provides the entire velocity distribution in a specified image region through estimation of the full power spectral density, and visualizes its evolution over time in a spectrogram \cite{welch1967use}. Spectral Doppler methods are relevant for e.g. detecting turbulent flow in stenotic arteries or across heart valves. Besides assessing blood flow, Doppler processing also finds applications in measurement of tissue velocities (tissue Doppler), e.g. for assessment of myocardial strain.

\subsection{Advanced applications}

\noindent In addition to B-mode, M-mode, and Doppler scanning, ultrasound data is used in a number of advanced applications. For instance, \textit{Elastography} methods aim at measuring mechanical parameters related to tissue elasticity, and rely on analysis of displacements following some form of imposed stress. Stress may be delivered manually (through gentle pushing), naturally (e.g in the myocardium of a beating heart) or acoustically, as done in acoustic radiation force impule imaging (ARFI) \cite{nightingale2011acoustic}. Alternatively, the speed of laterally traveling shear waves induced by an acoustic push-pulse can be measured, with this speed being directly related to the shear modulus \cite{bercoff2004supersonic}. Shear wave elasticity imaging (SWEI) also permits measurement of tissue viscosity in addition to stiffness through assessment of wave dispersion \cite{van2017viscoelasticity}. All the above methods rely on adequate measurement of local tissue velocity or displacement through some form of tissue Doppler processing. 

While Doppler methods enable estimation of blood flow, detection of low-velocity microvascular flow is challenging since its Doppler spectrum overlaps with that of the strong tissue clutter. \textit{Contrast-enhanced ultrasound} (CEUS) permits visualization and characterization of microvascular perfusion through the use of gas-filled microbubbles \cite{goldberg1994ultrasound,van2017ultrasound}. These intravascular bubbles are sized similarly to red blood cells, reaching the smallest capillaries in the vascular net, and exhibit a particular nonlinear response when insonified. The latter is specifically exploited in contrast-enhanced imaging schemes, which aim at isolating this nonlinear response through dedicated pulse sequences. Unfortunately, this does not lead to complete tissue suppression, since tissue itself also generates harmonics \cite{hamilton1998nonlinear}. Thus, clutter rejection algorithms are becoming increasingly popular, in particular when used in conjunction with ultrafast imaging \cite{desailly2016contrast}. 

Recent developments also leverage the microbubbles used in CEUS to yield \textit{super-resolution imaging} \cite{viessmann2013acoustic,oˈreilly2013super,bar2017fast,bar2018sushi}. Ultrasound localization microscopy (ULM) is a particularly popular approach to achieve this \cite{errico2015ultrafast}. ULM methods rely on adequate detection, isolation and localization of the microbubbles, typically achieved through precisely tuned tissue clutter suppression algorithms and by posing strong constraints on the allowable concentrations. We will further elaborate on this approach and its limitations in Section~IV, where we discuss a dedicated deep learning solution for super resolution ultraound that aims at addressing some of these disadvantages.

\section{Deep learning for (Front-end) \\ ultrasound processing}
\label{sec:deepfrontend}
\noindent The effectiveness of ultrasound imaging and its applications is dictated by adequate front-end beamforming, compression, signal extraction (e.g. clutter suppression) and velocity estimation. In this section we demonstrate how neural networks, being universal function approximators \cite{hornik1989multilayer}, can learn to act as powerful artificial agents and signal processors across the imaging chain to improve resolution and contrast, adequately suppress clutter, and enhance spectral estimation. We here refer to artificial agents \cite{mnih2015human} whenever these learned networks impact the processing chain by actively and adaptively changing the settings or parameters of a particular processor depending on the context.

Deep learning is the process of learning a hierarchy of parameterized nonlinear transformations (or layers) such that it performs a desired function. These elementary nonlinear transformations in a deep network can take many forms and may embed structural priors. A popular example of the latter is the translational invariance in images that is exploited by convolutional neural networks, but we will see that in fact many other structural priors can be exploited.

The methods proposed throughout this work are both model-based and learn from data. We complement this approach with a-priori knowledge on signal structure, to develop deep learning models that are both effective and data-efficient, i.e. `fast learners'. An overview is given in Fig.~\ref{fig:overview}. We assume that the reader is familiar with the basics of (deep) neural networks. For a general introduction to deep learning, we refer the reader to \cite{goodfellow2016deep}.

\subsection{Beamforming}
\label{sec:deep_beamforming}
%- Adaptive beamforming, exploiting the model structure of a minimum variance estimator (and its improvement over general-purpose models)
\subsubsection{Deep neural networks as beamformers}
The low complexity of delay-and-sum beamforming has made it the industry standard and commonplace for real-time ultrasound beamforming. There are however a number of factors that cause deteriorated reconstruction quality of this naive spatial filtering strategy. First, the channel delays for time-of-flight correction are based on the geometry of the scene and assume a constant speed of sound across the medium. As a consequence, variations in speed of sound and resulting aberrations impair proper alignment of echoes stemming from the same scatterer \cite{mallart1992sound}. Second, the a-priori determined channel weighting (apodization) of pseudo-aligned echoes before summation requires a trade-off between main-lobe width (resolution) and side-lobe level (leakage) \cite{van2004optimum}. 

Delay-and-sum beamformers are typically hand-tailored based on knowledge of the array geometry and medium properties, often including specifically designed array apodization schemes that may vary across imaging depth. Interestingly, it is possible to learn the delays and apodizations from paired channel-image data through gradient-descent by dedicated ``delay layers'' \cite{vedula2018learning}.  To show this, unfocused channel data was obtained from echocardiography of six patients for both single-line and multi-line acquisitions. While the latter allows for increased frame rates, it leads to deteriorated image quality when applying standard delay-and-sum beamforming. The authors therefore propose to train a more appropriate delay-and-sum beamforming chain that takes multi-line channel data as an input, and produces beamformed images that are as close as possible to those obtained from single-line-acquisitions, minimizing their $\ell_1$ distance. Since the introduced delay and apodization layers are differentiable, efficient learning is enabled through backpropagation. Although such an approach potentially enables discovery of a more optimal set of parameters dedicated to each application, the fundamental problem of having a-priori-determined static delays and weights remains.

Several other data-driven beamforming methods have recently been proposed. In contrast to \cite{vedula2018learning}, these are mostly based on ``general-purpose'' deep neural networks, such as stacked autoencoders \cite{perdios2017deep}, encoder-decoder architectures \cite{simson2019end}, and fully-convolutional networks that map pre-delayed channel data to beamformed outputs \cite{khan2019universal}. In the latter, a 29-layer convolutional network was applied to a 3D stack of array response vectors for all lateral positions and a set of depths, to yield a beamformed in-phase and quadrature output for those lateral positions and depths. Others exploit neural networks to process channel data in the Fourier domain \cite{luchies2018deep}. To that end, axially gated sections of pre-delayed channel data first undergo discrete Fourier-transformation. For each frequency bin, the array responses are then processed by a separate fully connected network. The frequency spectra are subsequently inverse Fourier-transformed and summed across the array to yield a beamformed radiofrequency signal associated to that particular axial location. The networks were specifically trained to suppress off-axis responses (outside the first nulls of the beam) from simulations of ultrasound channel data for point targets. 

Beyond beamforming for suppression of off-axis scattering, the authors in \cite{hyun2019beamforming} propose deep convolutional neural networks for joint beamforming and speckle reduction. Rather than applying the latter as a post-processing technique, it is embedded in the beamforming process itself, permitting exploitation of both channel and phase information that is otherwise irreversibly lost. The network was designed to accept 16 beamformed subaperture radio frequency (RF) signals as an input, and outputs speckle-reduced B-mode images.  The final beamformed images exhibit comparable speckle-reduction as post-processed delay-and-sum images using the optimized Bayesian nonlocal means algorithm \cite{coupe2009nonlocal}, yet at an improved resolution. Additional applications of deep learning in this context include removal of artifacts in time-delayed and phase-rotated element-wise I/Q data in multi-line acquisitions for high-frame-rate imaging \cite{senouf2018high}, and synthesizing multi-focus images from single-focus images through generative adversarial networks \cite{Goudarzi2019}. In \cite{nair2019generative}, such generative adversarial networks were used for joint beamforming and segmentation of cyst phantoms from unfocused RF channel data acquired after a single plane-wave transmission.

While the flexibility and capacity of very deep neural networks in principle allows for learning context-adaptive beamforming schemes, such highly overparameterized networks notoriously rely on vast RF channel data to yield robust inference under a wide range of conditions. Moreover, large networks have a large memory footprint, complicating resource-limited implementations.  \\ %something on adversarial examples and non-robustness of large neural networks to new data?
% Add paragraph of paper on "universal deep beamformer for variable rate ultrasound"

\subsubsection{Leveraging model-based algorithms}
One approach to constraining the solution space while explicitly embedding adaptivity is to borrow concepts from model-based adaptive beamforming methods. These techniques steer away from the fixed-weight presumption and calculate an array apodization depending on the measured signal statistics. In the case of pixel-based reconstruction, apodization weights can be adaptively optimized per pixel. A popular adaptive beamforming method is the minimum variance distortionless response (MVDR), or Capon, beamformer, where optimal weights are defined as those that minimize signal variance/power, while maintaining distortionless response of the beamformer in the desired source direction. This amounts to solving: 
\begin{equation} 
\label{eqn:beamforming_capon}
\begin{aligned}
\hat{\mathbf{w}} = \underset{\mathbf{w}}{\arg\min} \;&\mathbf{w}^H \mathbf{R}_x \mathbf{w} \\ 
\text{s.t. } &\mathbf{w}^H \mathbf{a}=1,
\end{aligned}
\end{equation}
where $\mathbf{R}_x$ denotes the covariance matrix calculated over the receiving array elements and $\mathbf{a}$ is a steering vector. When receive signals are already time-of-flight corrected, $\mathbf{a}$ is a unity vector.

\begin{figure*}[t!]
	\centering
	\includegraphics[trim=0cm 8.2cm 10.8cm 0cm, clip=true,width=500px]{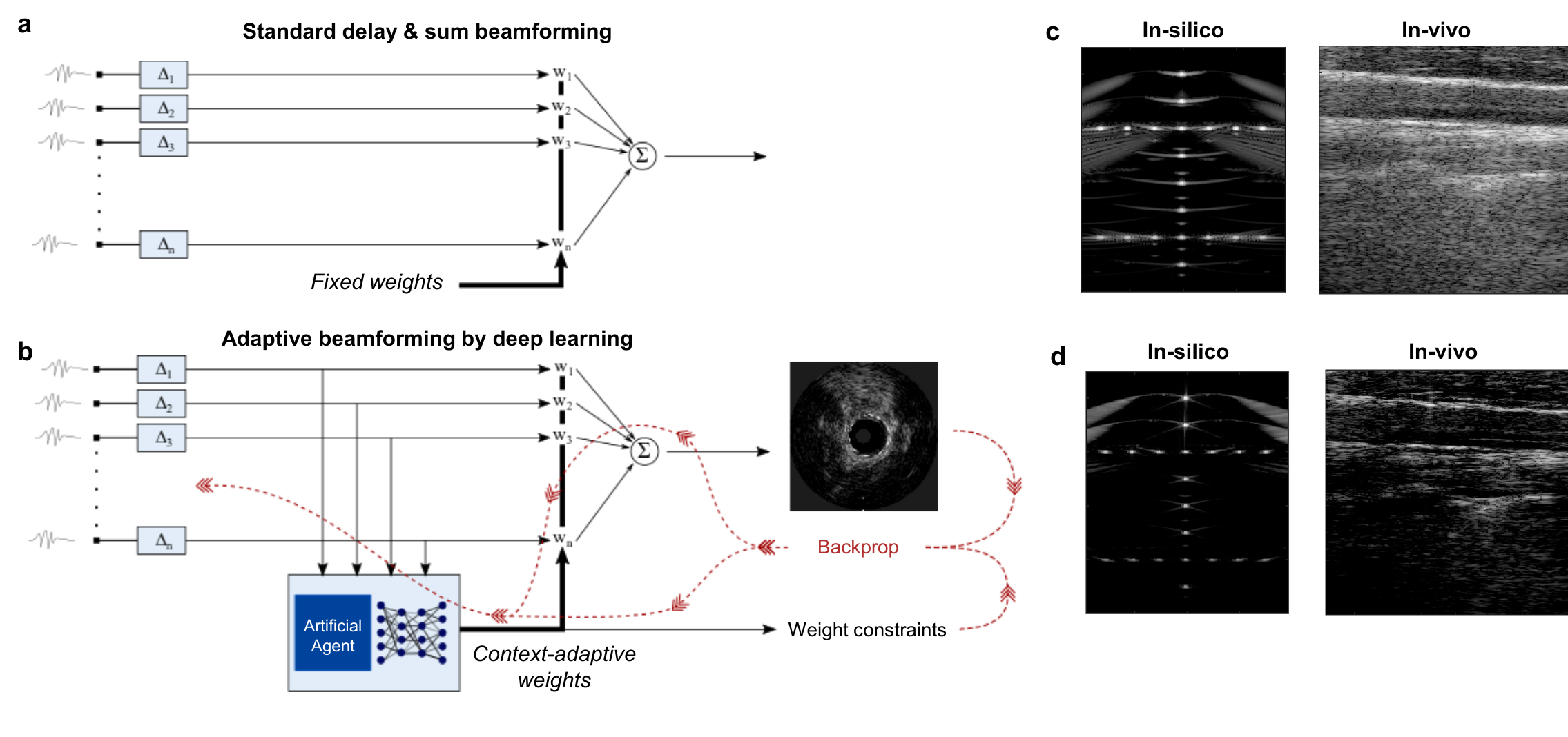}
	\caption{(a) Flow charts of standard delay-and-sum beamforming using fixed apodization weights, and (b) adaptive beamforming by deep learning \cite{Luijten2019beamforming}, along with (c,d) illustrative reconstructed images (\textit{in-silico} and \textit{in-vivo}) for both methods, respectively. Adaptive beamforming by deep learning achieves notably better contrast and resolution and generalizes very well to unseen datasets.}
	\label{fig:Beamforming}
\end{figure*}

Solving \eqref{eqn:beamforming_capon} involves the inversion of $\mathbf{R}_x$, whose computational complexity grows cubically with the number of array elements \cite{boyd2004convex}. To improve stability, it is often combined with subspace selection through eigendecomposition, further increasing the computational burden. Another problem is the accurate estimation of $\mathbf{R}_x$, typically requiring some form of averaging across sub-arrays and the fast- and slow-time scales. While this implementation of MVDR beamforming is impractical for typical ultrasound arrays (e.g $256$ elements) or matrix-transducers (e.g $64\times64$ elements), it does provide a framework in which deep neural networks can be leveraged efficiently and effectively.

Instead of attempting to replace the beamforming process entirely, a neural network can be used specifically to act as an artificial agent that calculates the optimal apodization weights $\mathbf{w}$ for each pixel, given the received pre-delayed channel signals at the array. By only replacing this bottleneck component in the MVDR beamformer, and constraining the problem further by enforcing close-to-distortionless response during training (i.e. $\Sigma_iw_i\approx1$), this solution is highly data-efficient, interpretable, and has the ability to learn powerful models from only few images \cite{Luijten2019beamforming}. 

The neural network proposed in \cite{Luijten2019beamforming} is compact, consisting of four fully connected layers comprising 128 nodes for the input and output layers, and 32 nodes for the hidden layers. This dimensionality reduction enforces compact representation of the data, mitigating the impact of noise. Between every fully connected layer, dropout is applied with a probability of 0.2.
The input of the network is the pre-delayed (focused) array response for a particular pixel (i.e. a vector of length $N$, with $N$ being the number of array elements), and its outputs are the corresponding array apodizations $\mathbf{w}$. This apodization is subsequently applied to the network inputs to yield a beamformed pixel. Since pixels are processed independently by the network, a large amount of training data is available per acquisition. Inference is fast and real-time rates are achievable on a GPU-accelerated system. For an array of 128 elements, adaptive calculation of a set of apodization weights through MVDR requires $>N^3 (=2,097,152)$ floating point operations (FLOPS), while the deep-learning architecture only requires 74656 FLOPS {\cite{Luijten2019beamforming}}, leading to a more than $400\times$ speed-up in reconstruction time. Additional details regarding the adopted network and training strategy are given in Section~\ref{sec:deepfrontend}-\ref{sec:design_training}.

Fig.~\ref{fig:Beamforming} exemplifies the effectiveness of this approach on plane-wave ultrasound acquisitions obtained using a linear array transducer. Compared to standard delay-and-sum, adaptive beamforming with a deep network serving as an artificial agent visually provides reduced clutter and enhanced tissue contrast. Quantitatively it yields a slightly elevated contrast-to-noise ratio (10.96~dB vs 11.48~dB), along with significantly improved resolution (0.43~mm vs 0.34~mm, and 0.85~mm vs 0.70~mm in the axial and lateral directions, respectively).

Interestingly, the neural network exhibits increased stability and robustness compared to the MVDR weight estimator. This can be attributed to its small bottleneck latent space, enforcing apodization weight realizations that are represented in a compact basis.
\\

\begin{figure*}[t!]
	\centering
	\includegraphics[trim=0cm 9.8cm 10.5cm 0cm, clip=true,width=500px]{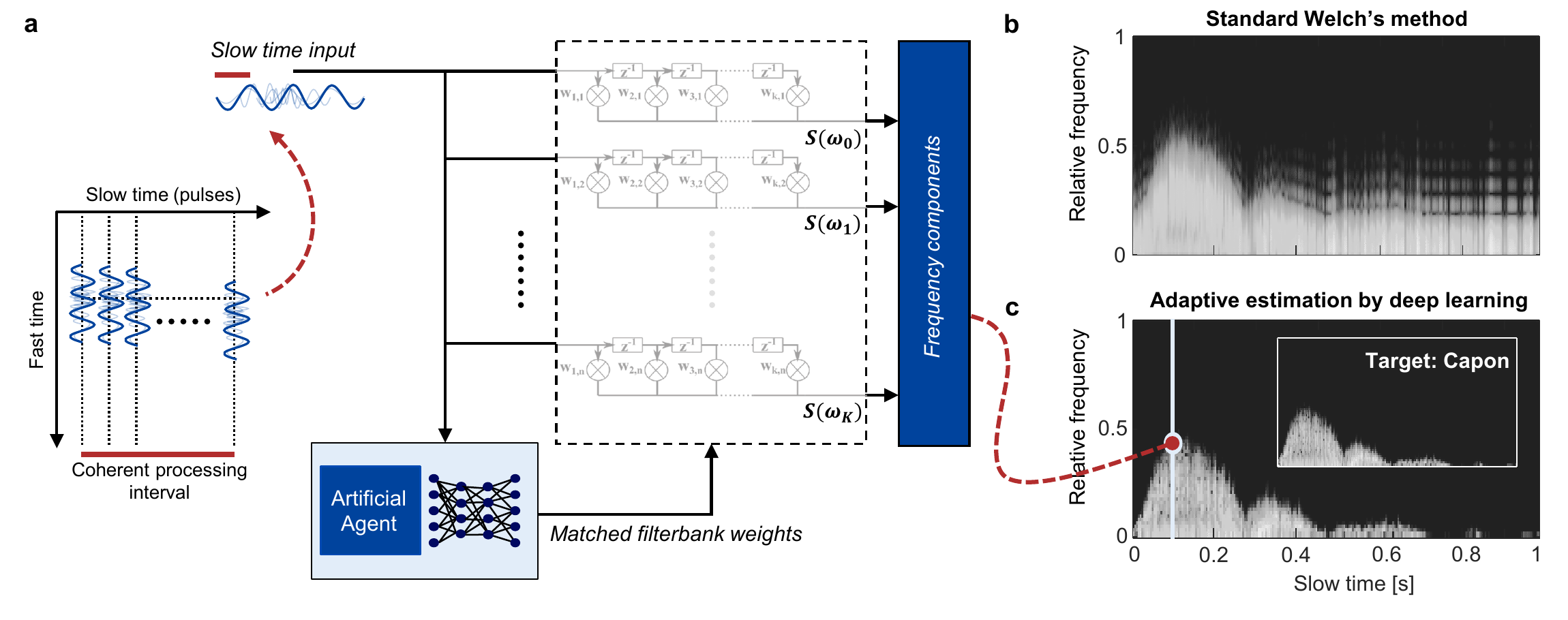}
	\caption{Adaptive spectral Doppler processing using deep learning, displaying (a) an illustrative overview of the method, comprising an artificial agent that adaptively sets the optimal matched filterbank weights according to the input data, and (b,c) spectral estimates using Welch's method and the deep learning approach, respectively. The input was phantom data for the arteria femoralis, with spectra estimated from a coherent processing interval of 64 slow-time samples. }
	\label{fig:spectralDoppler}
\end{figure*}

\subsubsection{Design and training considerations}
\label{sec:design_training}
The large dynamic range and modulated nature of radio-frequency ultrasound channel data motivates the use of specific nonlinear activation functions. While rectified linear units (ReLUs) are typically used in image processing, popular for their sparsifying nature and ability to avoid vanishing gradients due to their positive unbounded output, it inherently causes many `dying nodes' (neurons that do no longer update since their gradient is zero) for ultrasound channel data, as a ReLU does not preserve (the abundant) negative values. To circumvent this, a hyperbolic tangent function could be used. Unfortunately, the large dynamic range of ultrasound signals makes it difficult to be in the `sweet spot', where gradients are sufficiently large, thereby avoiding vanishing gradients during backpropagation across multiple layers.

A powerful alternative that is by nature unbounded and preserves both positive and negative values is the class of concatenated rectified linear units \cite{shang2016understanding}. A particular case is the \textit{anti-rectifier} function:
\begin{equation}
    \label{eqn:antirectifier}
    f(\mathbf{x}) = \left(\begin{matrix} \left[ \frac{\mathbf{x}-\bar{\mathbf{x}}}{\left||\mathbf{x}-\bar{\mathbf{x}}\right||_2}\right]_{+} \\ \left[-\frac{\mathbf{x}-\bar{\mathbf{x}}}{\left||\mathbf{x}-\bar{\mathbf{x}}\right||_2}\right]_{+}  \end{matrix}\right),
\end{equation}
where $[\cdot]_+ = \textrm{max}(\cdot,0)$ is the positive part operator, $\mathbf{x}$ is a vector containing the linear responses of all neurons (before activation) at a particular layer, and $\bar{\mathbf{x}}$ is its mean value across all those neurons. The anti-rectifier does not suffer from vanishing gradients, nor does it lead to dying nodes for negative values, yet provides the nonlinearity that facilitates learning complex models and representations. This dynamic-range preserving activation scheme is therefore well-suited for processing radio-frequency or IQ-demodulated ultrasound channel data, and is also used for the results presented in Fig.~\ref{fig:Beamforming}. These advantages come at the cost of a higher computational complexity compared to a standard ReLU activation.

When training a neural-network-based ultrasound beamforming algorithm, it is important to consider the impact of subsequent signal transformations in the processing chain. In particular, envelope-detected beamformed signals typically undergo significant dynamic range compression (e.g. through a logarithmic transformation) to project the high dynamic range of backscattered ultrasound signals onto the limited dynamic range of a display, and allow for improved interpretation and diagnostics. To incorporate this aspect in the neural network's training loss, beamforming errors can be transformed to attain a mean squared logarithmic error:
\begin{align}
    \label{eqn:logerror}
    \mathcal{L} =& \left\Vert\textrm{log}_{10}([\hat{\mathbf{y}}]_+) - \textrm{log}_{10}([\mathbf{y}]_+)\right\vert_2^2 \nonumber \\ &+\left\vert\textrm{log}_{10}([-\hat{\mathbf{y}}]_+) - \textrm{log}_{10}([-\mathbf{y}]_+)\right\Vert_2^2,
\end{align}
where $\hat{\mathbf{y}}$ is a vector containing the neural-network-based prediction of the beamformed responses for all pixels, and $\mathbf{y}$ contains the target beamformed signals. For our model-based adaptive beamforming solution \cite{Luijten2019beamforming}, $\mathbf{y}$ contains the MVDR beamformer outputs for each pixel, and $\mathbf{y}$ is the corresponding set of pixel responses after application of the apodization weights calculated by the neural network.

\subsection{Adaptive spectral estimation for spectral Doppler}
\label{sec:spectraldoppler}
\noindent As mentioned in Section~II, beamformed ultrasound signals are not only used to visualize anatomical information in B-mode, they also permit the extraction of velocities by processing subsequent frames across slow-time. 

Spectral Doppler ultrasound enables measurement of blood (and tissue) velocity distributions through the generation of a Doppler spectrogram from slow-time data sequences, i.e. a series of subsequent pulse-echo snapshots. In commercial systems, spectra are estimated using Fourier-transform-based periodogram methods, e.g. the standard Welch approach. Such techniques however require long observation windows (denoted as `coherent processing intervals') to achieve high spectral resolution and mitigate spectral leakage. This deteriorates the temporal resolution.

Data-adaptive spectral estimators alleviate the strong time-frequency resolution tradeoff, providing superior spectral estimates and resolution for a given temporal resolution \cite{gran2009adaptive}. The latter is determined by the coherent processing interval, which is in turn defined by the pulse repetition frequency and the number of slow-time snapshots required for a spectral estimate. Adaptive approaches steer away from the standard periodogram methods, and rely on content-matched filterbanks. The filter coefficients for each frequency of interest $\omega$ are adaptively tuned to e.g. minimize signal energy while being constrained to unity frequency response. This Capon spectral estimator is given by solving \cite{gran2009adaptive}:
\begin{equation} 
\label{eqn:spectral_capon}
\begin{aligned}
\hat{\mathbf{w}}_\omega = \underset{\mathbf{w}_\omega}{\arg\min} \;&\mathbf{w}_\omega^H \mathbf{R}_y \mathbf{w}_\omega \\ 
\text{s.t. } &\mathbf{w}_\omega^H \mathbf{e}_\omega=1,
\end{aligned}
\end{equation}
where $\mathbf{R}_y$ is the covariance matrix of the (slow-time) input signal vector $\mathbf{y}$, and $\mathbf{e}_\omega$ is the corresponding Fourier vector. While this adaptive spectral estimator indeed improves upon standard approaches and significantly lowers the required observation window while gaining spectral fidelity, it unfortunately suffers from high computational complexity stemming from the need for inversion of the signal covariance matrix.

As for the MVDR beamformer (Section III-\ref{sec:deep_beamforming}), we here demonstrate that neural networks can also be exploited to provide fast estimators for the optimal matched filter coefficients, acting as an artificial agent. An overview of this approach is given in Fig.~\ref{fig:spectralDoppler}, for a pulsed-wave phantom data for the arteria femoralis \cite{jensen1996estimation}. The neural network takes a beamformed slow-time RF signal as input, and outputs a set of filter coefficients for each filter in the filterbank. The slow-time input signal is then passed through this filterbank to attain a spectral estimate. The neural network is trained by minimizing the mean squared logarithmic error \eqref{eqn:logerror} between the resulting spectrum and the output spectrum of the high-quality adaptive Capon spectral estimator. It comprised 128 4-layer fully-connected subnetworks, each of those predicting the coefficients for one of the 128 filters in the filterbank. The optimization problem is then regularized by penalizing deviations from unity frequency response \eqref{eqn:spectral_capon}. The length of the slow-time observation window was only 64 samples, taken from a single depth sample. Compared to Welch's periodogram-based method, adaptive spectral estimation by deep learning achieves far less spectral leakage, and higher spectral resolution (Fig. ~\ref{fig:spectralDoppler}b and c).

Training the artificial agent is subject to similar considerations outlined in Section~III-A4. First, slow-time input samples have a large dynamic range such that a non-saturating activation scheme is preferred \eqref{eqn:antirectifier}. Second, Doppler spectra are typically presented in decibels, advocating for the use of a log-transformed training loss as in \eqref{eqn:logerror}. Third, training is regularized by adding an additional loss to penalize predicted filterbanks that deviate from unity frequency response.

The above approach is designed to processes uniformly sampled slow-time signals. In practice, there is a desire to expand these techniques to estimators that have the ability to cope with `gaps', or even sparsely sampled signals, since spectral Doppler processing is typically interleaved with B-mode imaging for navigation purposes (Duplex mode). To that end, extensions of  data-adaptive estimators for periodically gapped data \cite{liu2009periodically}, and recovery for nested slow-time sampling \cite{cohen2018sparse} can be used.

\subsection{Compressive encodings for tissue Doppler}
\label{sec:DopplerNet}

\noindent From a hardware perspective, a significant challenge for the design of ultrasound devices and transducers is coping with the limited cable bandwidth and related connectivity constraints \cite{rashid2016front}. This is particularly troublesome for catheter transducers used in interventional applications (e.g. intra-vascular ultrasound or intra-cardiac echography), where data needs to pass through a highly restricted number of cables. While this is less of a concern for transducers with only few elements, the number of transducer elements have expanded greatly in recent devices to facilitate high-resolution 2D or 3D imaging \cite{dausch2014vivo}. Beyond the limited capacity of miniature devices, (future) wireless transducers will pose similar constraints on data rates \cite{bar2015towards}. Today, front-end connectivity and bandwidth challenges are addressed through e.g. application-specific integrated circuits that perform microbeamforming \cite{wildes20164} or simple summation of the receive signals across neighbouring elements \cite{bera2017dual} to compress the full channel data into a manageable amount, and multiplexing of the receive signals. This inherently entails information loss, and typically leads to reduced image quality. %Alternatively, dedicated compressive sampling strategies can be used [citations to Xampling work Yonina].  

Instead of Nyquist-rate sampling of pre-beamformed and multiplexed channel data, compressive sub-Nyquist sampling methods permit reduced-rate imaging without sacrificing quality \cite{chernyakova2014fourier,chernyakova2018fourier}. After (reduced-rate) digitization, additional compression may be achieved through neural networks that serve as application-specific encoders. Advances in low-power neural edge computing may permit placing such a trained encoder at the probe, further alleviating probe-scanner communication, and a subsequent high-end decoder at the remote processor \cite{teerapittayanon2017distributed}. 

%Deep neural networks have been shown capable of learning powerful compressive encodings and structured signal recovery \cite{mousavi2015deep}. Deep \textit{autoencoders} exploit compressive nonlinear encoding layers to condense the input signals into a compact latent representation. A jointly trained hierarchy of decoding layers then decodes this representation to provide recovery, enabling efficient image compression that is up to par with that of JPEG2000 \cite{theis2017lossy}. Unlike the latter, these deep autoencoders can learn compressive representations that are strongly optimized for specific content. In addition, they can serve as competent denoisers that can be trained in an unsupervised fashion, e.g. through leveraging Stein's unbiased risk estimate \cite{metzler2018unsupervised}. 

\begin{figure*}[t!]
	\centering
	\includegraphics[trim=0cm 6.5cm 12cm 0cm, clip=true,width=480px]{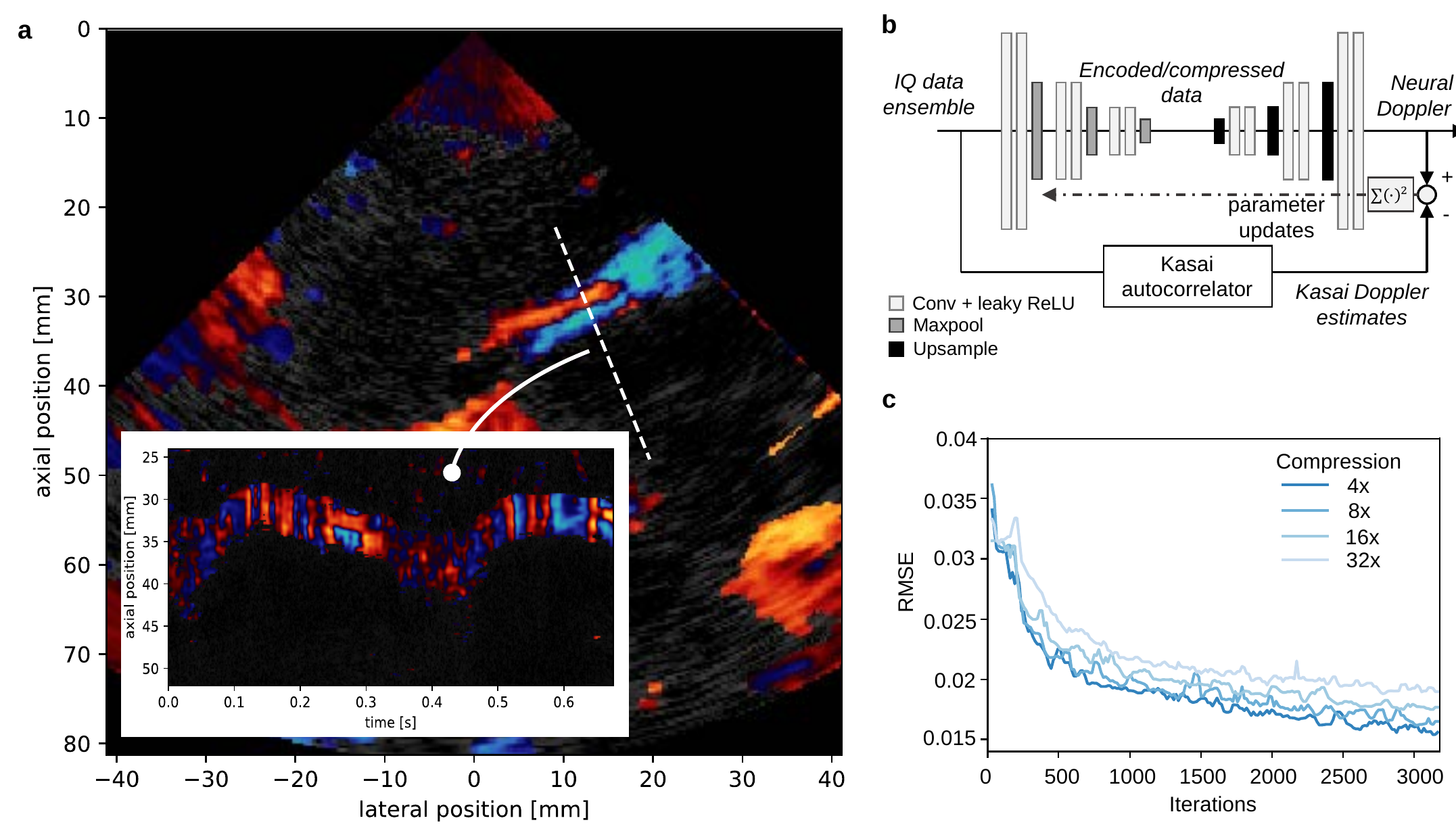}
	\caption{(a) Tissue Doppler processing using a deep encoder-decoder network for an illustrative intra-cardiac ultrasound application \cite{van2018learning}, displaying the wall between the right atrium and the aorta. (b) The deep network architecture is designed to encode input IQ data into a compressed latent space via a series of convolutional layers and spatial (max) pooling operations, while maintaining the functionality and performance of a typical Doppler processor (Kasai autocorrelator \cite{loupas1995axial}) using full uncompressed IQ data. (c) Convergence of the network parameters during training, showing the relative root-mean-squared-errors (RMSE) on a test dataset for four data compression factors. }
	\label{fig:colorDoppler}
\end{figure*}

Instead of aiming at decoding the full input signals from the encoded representation, one can also envisage decoding only a specific signal or source that is to be extracted from the input. This may enable stronger compression during encoding whenever this component has a more restricted entropy than the full signal. In ultrasound imaging, such signal-extracting compressive deep encoder-decoders can e.g. be used for velocity estimation in colour Doppler \cite{van2018learning}. Fig.~\ref{fig:colorDoppler} shows how these networks enable decoding of tissue Doppler signals from encoded IQ-demodulated input data acquired in an \textit{in-vivo} open-chest experiment of a porcine model, using intra-cardiac diverging-wave imaging in the right atrium at a frame rate of 474 Hz.

Here the encoding neural network comprised a series of three identical blocks, each composed of two subsequent convolutional layers across fast- and slow-time, followed by an aggregation of this processing through spatial downsampling (max pooling). The decoder had a similar, mirrored, architecture. The degree of IQ data compression achieved by the encoder can be changed by varying the number of channels (in the context of image processing often referred to as feature maps) at the latent layer. The encoder and decoder network parameters can then be learnt by mimicking the phase (and therewith, velocity) estimates obtained using the well-know Kasai autocorrelator on the full input data (see Fig. \ref{fig:colorDoppler}b). Interestingly, IQ compression rates as high as 32 can be achieved (see Fig. {\ref{fig:colorDoppler}}c), while retaining reasonable Doppler signal quality, yielding a relative phase root-mean-squared-error of approximately $0.02$. These errors drop when requiring lower compression rates. Higher compression rates lead to an increased degree of spatial consistency, displaying fewer spurious variations which could not be represented in the compact latent encoding.  

The design of traditional Doppler estimators involves careful optimization of the slow- and fast-time range gates across which the estimation is performed, amounting to a trade-off between the estimation quality and spatiotemporal resolution \cite{loupas1995axial}. For many practical applications, the optimal settings not only vary across measurements and desired clinical objectives, but also within a single measurement. In contrast, a convolutional encoder-decoder network can learn to determine the effective spatiotemporal support of the given input data required for adequate Doppler encoding and prediction. 

%Using GPU acceleration encoding and decoding is performed at a rate of about 60 complete Doppler frame reconstructions per second. 

%\subsection{Learning sampling schemes and multiplexing}
%- Color Doppler by diluting slow-time (hopefully in the near future also across the channels).

\subsection{Unfolding Robust PCA for clutter suppression}
\label{sec:RPCA}

\noindent An important ultrasound-based modality is contrast-enhanced ultrasound (CEUS) \cite{furlow2009contrast}, which allows the detection and visualization of small blood vessels. In particular, CEUS is used for imaging perfusion at the capillary level \cite{lassau2007dynamic,hudson2015dynamic}, and for estimating different properties of the blood such as relative volume, velocity, shape and density. These physical parameters are related to different clinical conditions, including cancer \cite{opacic2018motion}.

The main idea behind CEUS is the use of encapsulated gas microbubbles, serving as ultrasound contrast agents (UCAs), which are injected intravenously and can flow throughout the vascular system due to their small size \cite{de1991principles}. To visualize them, strong clutter signals originating from stationary or slowly moving tissues must be removed as they introduce significant artifacts in the resulting images \cite{bjaerum2002clutter}. The latter poses a major challenge in ultrasonic vascular imaging and various methods have been proposed to address it.  In \cite{thomas1994improved}, an high-pass filtering approach was presented to remove tissue signals using finite impulse response (FIR) or infinite impulse response (IIR) filters. However, this approach is prone to failure in the presence of fast tissue motion. An alternative strategy is second harmonic imaging \cite{frinking2000ultrasound} which exploits the non-linear response of the UCAs to separate them from the tissue. This technique, however, does not remove the tissue completely as it also exhibits a nonlinear response.

One of the most popular approaches for clutter suppression is spatio-temporal filtering based on the singular value decomposition (SVD). This strategy has led to various techniques for clutter removal \cite{bjaerum2002clutter,yu2010eigen,mauldin2010complex,mauldin2011singular,gallippi2003bss,lovstakken2006real,kruse2002new,Errico2015,song2017ultrasound,chee2018receiver,kim2018multidimensional}. SVD filtering includes collecting a series of consecutive frames, stacking them as vectors in a matrix, performing SVD of the matrix and removing the largest singular values, assumed to be related to the tissue. Hence, a crucial step in SVD filtering is determining an appropriate threshold which discriminates between tissue related and blood related singular values.
However, the exact setting of this threshold is difficult to determine and may vary dramatically between different scans and subjects, leading to significant defects in the constructed images.

\begin{figure*}[t!]
	\centering
	\includegraphics[trim=0cm 0cm .5cm 0cm, clip=true,width=500px]{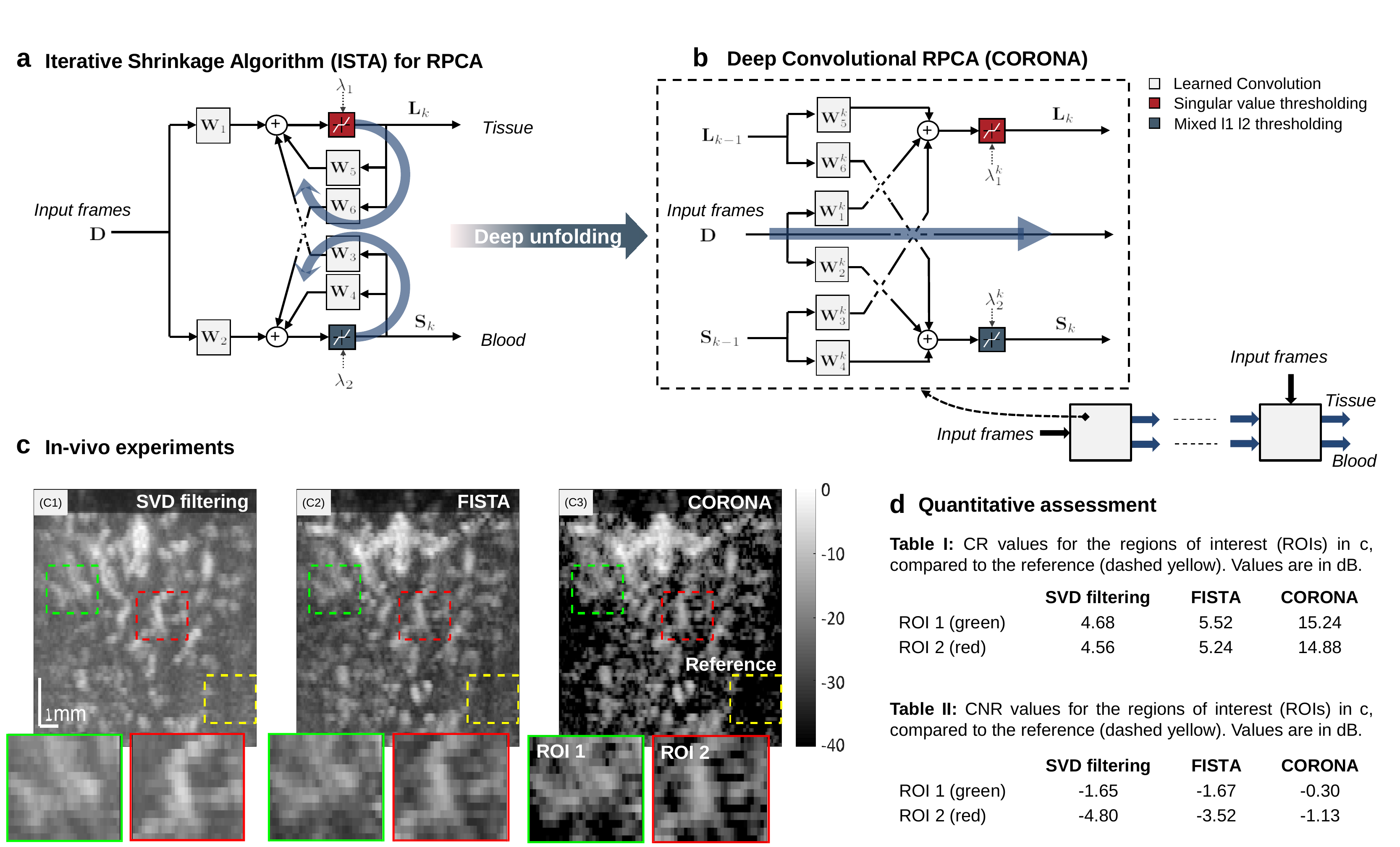}
	\caption{(a) ISTA diagram for solving RPCA and (b) a diagram of a single layer of CORONA \cite{solomon2018deep}. (c) Qualitative assessment of clutter removal performed by SVD filtering, FISTA and CORONA, shown in panels $\bf c_1-c_3$ respectively. Below each panel, we present enlarged views of selected areas, indicated by the green and red rectangles. (d) Quantitative assessment of clutter removal performed by the mentioned methods.}
	\label{fig:CORONA}
\end{figure*}

To overcome these limitations, in \cite{solomon2018deep,ashikuzzaman2019suppressing,bayat2018concurrent}, the task of clutter removal was formulated as a convex optimization problem by leveraging a low-rank-and-sparse decomposition. The authors of \cite{solomon2018deep} then proposed an efficient deep learning solution to this convex optimization problem through an algorithm-unfolding strategy \cite{li2019algorithm}. To enable explicit embedding of signal structure in the resulting network architecture, the following model for the signal after beamforming was proposed.

Denote the received beamformed signal at snapshot time $t$ by $D(x,z,t)$, where $(x,z)$ are image coordinates. Then we may write:
\begin{equation}
    D(x,z,t) = L(x,z,t)+S(x,z,t),
\end{equation}
where the term $L(x,z,t)$ represents the tissue and $S(x,z,t)$ is the signal stemming from the blood.
Similar to SVD filtering, a series of consecutive snapshots ($t=1,...,T$) is acquired and stacked as vectors into a matrix, leading to the matrix model:
\begin{equation}
    \bf D = L+S.
\end{equation}
The tissue exhibits high spatio-temporal coherence, hence, the matrix $\bf L$ is assumed to be low rank. The matrix $\bf S$ is considered to be sparse, since small blood vessels sparsely populate the image plane. 

These assumptions on the rank of $\bf L$ and the sparsity in $\bf S$ enable formulation of the task of clutter suppression as a \textit{robust principle component analysis} (RPCA) problem \cite{otazo2015low}:
\begin{equation}
\label{Eq:minprob}
\min_{\bf L,S}\frac{1}{2}||{\bf D - (L +S)}||_F^2+\lambda_1||{\bf L}||_{*}+\lambda_2|| {\bf S}||_{1,2},
\end{equation}
where $\lambda_1$ and $\lambda_2$ are threshold parameters. The symbol $||\cdot||_*$ stands for the nuclear norm, which sums the singular values of $\bf L$. The term $||\cdot||_{1,2}$ is the mixed $l_{1,2}$ norm \cite{bar2018sushi,solomon2019sparcom}, which promotes sparsity of the blood vessels along with consistency of their locations over consecutive frames. RPCA is widely used in the area of computer vision, and can be solved iteratively using the fast iterative shrinkage/soft-thresholding algorithm (FISTA) \cite{beck2009fast}, leading to the following update rules
\begin{align}
    \begin{split}
        &{\bf L}^{k+1}=\mathcal{ST}_{\lambda_1/2}\left(\frac{1}{2}{\bf L}^k-{\bf S}^k+
        {\bf D}\right), \\
        &{\bf S}^{k+1}=\mathcal{MT}_{\lambda_2/2}\left(\frac{1}{2}{\bf S}^k-{\bf L}^k+
        {\bf D}\right).
        \label{eq:fista}
    \end{split}
\end{align}
Here $\mathcal{MT}_{\alpha}({\bf X})$ is the mixed $\ell_{1,2}$ soft-thresholding operator which applies the function $\max(0,1-\frac{\alpha}{||{\bf x}||}){\bf x}$ on each row $\bf x$ of the input matrix $\bf X$. Assuming the input matrix is given by its SVD ${\bf X=U\Sigma V}^H$, the singular value thresholding (SVT) is defined as $\mathcal{ST}_{\alpha}({\bf X)=U\mathcal{S}_\alpha(\Sigma)V}^H$ where $\mathcal{S}_\alpha(x)=\max(0,x-\alpha)$ is applied point-wise on $\bf \Sigma$. A diagram of this iterative solution is given in Fig.~\ref{fig:CORONA}a. 

As shown in Fig.~\ref{fig:CORONA}c, the iterative solution \eqref{eq:fista} outperforms SVD filtering and leads to improved clutter suppression. However, it suffers from two major drawbacks. The threshold parameters $\lambda_1, \lambda_2$ need to be properly tuned as they have a significant impact on the final result. Moreover, depending on the dynamic range between the tissue and the blood, FISTA may require many iterations to converge, thus, making it impractical for real-time imaging. This motivates the pursuit of a solution with fixed complexity in which the threshold parameters are adjusted automatically. 

Such a fixed-complexity solution can be attained through \textit{unfolding} \cite{gregor2010learning,lecun2015deep}, in which a known iterative solution is unrolled as a feedforward neural network. In this case, the iterative solution is the FISTA algorithm (\ref{eq:fista}), which can be rewritten as
\begin{align}
    \begin{split}
        &{\bf L}^{k+1}=\mathcal{ST}_{\lambda_1/2}\left({\bf W}_1{\bf D}+{\bf W}_3{\bf S}^k+{\bf W}_5{\bf L}^k\right), \\
        &{\bf S}^{k+1}=\mathcal{MT}_{\lambda_2/2}\left({\bf W}_2{\bf D}+{\bf W}_4{\bf S}^k+{\bf W}_6{\bf L}^k\right).
    \end{split}
    \label{eq:towardsdeep}
\end{align}
Here ${\bf W}_1={\bf W}_2={\bf I}$, ${\bf W}_3={\bf W}_6=-{\bf I}$ and ${\bf W}_4={\bf W}_5=\frac{1}{2}{\bf I}$. From this, the deep multi-layer network takes a form in which the $k$th layer is given by
\begin{align}
    \begin{split}
        &{\bf L}^{k+1}=\mathcal{ST}_{\lambda_1^k}\left({\bf W}^k_1\ast{\bf D}+{\bf W}^k_3\ast{\bf S}^k+{\bf W}^k_5\ast{\bf L}^k\right), \\
        &{\bf S}^{k+1}=\mathcal{MT}_{\lambda_2^k}\left({\bf W}^k_2\ast{\bf D}+{\bf W}^k_4\ast{\bf S}^k+{\bf W}^k_6\ast{\bf L}^k\right).
    \end{split}
    \label{eq:corona}
\end{align}
In (\ref{eq:corona}), the matrices $\left({\bf W}_1^k,\cdots,{\bf W}_6^k\right)$ and the regularization parameters $\lambda_1^k$ and $\lambda_2^k$ differ from one layer to another and are learned during training. Moreover, $\left({\bf W}_1^k,\cdots,{\bf W}_6^k\right)$ were chosen to be convolution kernels where $\ast$ denotes the convolution operator. The latter facilitates spatial invariance along with a notable reduction in the number of learned parameters. This results in a CNN that is specifically tailored for solving RPCA, whose non-linearities are the soft-thresholding and SVT operators, and is termed Convolutional rObust pRincipal cOmpoNent Analysis (CORONA). A diagram of a single layer from CORONA is given in Fig.~\ref{fig:CORONA}b.

The training process of CORONA is performed by back-propagation in a supervised manner, leveraging both simulations, for which the true decomposition is known, and \textit{in-vivo} data for which the decomposition of FISTA \eqref{eq:fista} is considered as the ground truth. Moreover, data augmentation is performed and the training is done on 3D patches extracted from the input measurements.       
The loss function was chosen as the sum of mean squared errors (MSE)
\begin{equation*}
    {\bf E}(\theta)=\frac{1}{2N}\left(\sum_{i=1}^N ||{\bf S}_i-\hat{\bf S}_i(\theta)||_F^2+||{\bf L}_i-\hat{\bf L}_i(\theta)||_F^2\right)
\end{equation*}
where $\left\{{\bf S}_i,{\bf L}_i\right\}_{i=1}^N$ are the ground truth and  $\left\{\hat{\bf S}_i,\hat{\bf L}_i\right\}_{i=1}^N$ are the network's outputs. The learned parameters are denoted by ${\theta}=\left\{ {\bf W}_1^k,\cdots,{\bf W}_6^k,\lambda_1^k,\lambda_2^k\right\}_{k=1}^K$ where $K$ is the number of layers. Backpropagation through the SVD was done using PyTorch's Autograd function \cite{paszke2017automatic}. 

Fig.~\ref{fig:CORONA} shows how CORONA effectively suppresses clutter on contrast-enhanced ultrasound scans of two rat brains, outperforming SVD filtering and RPCA through FISTA \eqref{eq:fista}. The recovered CEUS (blood) signals are given in Fig.~\ref{fig:CORONA}c, including enlarged views of regions of interest. Visually judging, FISTA achieves moderately better contrast than SVD filtering, while CORONA outperforms both approaches by a large margin. For a quantitative comparison, the contrast-to-noise ratio (CNR) and contrast ratio (CR) were assessed, defined as 
\begin{equation*}
\label{eq:CNR}
\textrm{CNR}=\frac{|\mu_s-\mu_b|}{\sqrt{\sigma_{s}^{2}+\sigma_{b}^{2}}},\quad\;\textrm{CR}=\frac{\mu_s}{\mu_b},
\end{equation*}
where $\mu_s$ and $\sigma_{s}^{2}$ are the mean and variance of the regions of interest in Fig.~\ref{fig:CORONA}c, and $\mu_b$ and $\sigma_{b}^{2}$ are the mean and variance of the noisy reference area indicated by the yellow box. In both metrics, higher values imply higher contrast ratios, which suggests better noise suppression. FISTA obtained slightly better performance than SVD filtering (CR $\approx4.6$dB and $\approx5.4$~dB, respectively) and CORONA outperformed both (CR $\approx15$~dB). In most cases, the performance of CORONA was about an order of magnitude better than that of SVD. Thus, combining a model for the separation problem with a data-driven approach leads to improved separation of UCA and tissue signals, together with noise reduction as compared to the popular SVD approach.

The complexity of all three methods is governed by the singular-value decomposition which requires $O(MN^2)$ FLOPS for an $M\times N$ matrix, where $M\geq N$. However, FISTA may require thousands of iterations, i.e., thousands of such SVD operations. Hence, FISTA for RPCA is computationally significantly heavier than regular SVD-filtering. On the other hand, for CORONA, up to 10 layers were shown to be sufficient (i.e., up to 10 SVD operations), therewith offering a dramatic increase in performance at the expense of only a moderate increase in complexity. All three methods can benefit from using inexact decompositions that exhibit reduced computational load, such as the truncated SVD and randomized SVD.

% \begin{table}[h!]
%   \begin{center}
%     \caption{CR values for the selected green and red rectangles in (c) above, as compared with the yellow background rectangle in each corresponding panel. All values are in dB.}
%     \label{tab:table2}
%     \begin{tabular}{|c|c|c|s|}
%       \hline % <-- Toprule here
%       & \textbf{SVD Filtering} & \textbf{FISTA} & \textbf{CORONA}\\
%       \hline % <-- Midrule here
%       Green box & 4.68 & 5.52 & 15.24\\
%       Red box   & 4.56 & 5.24 & 14.88\\
%       \hline % <-- Bottomrule here
%     \end{tabular}
%   \end{center}
% \end{table}

% \begin{table}[h!]
%   \begin{center}
%     \caption{\small CNR values for the selected green and red rectangles of (c) above, as compared with the yellow background rectangle in each corresponding panel. All values are in dB.}
%     \label{tab:table1}
%     \begin{tabular}{|c|c|c|s|}
%       \hline % <-- Toprule here
%       & \textbf{SVD Filtering} & \textbf{FISTA} & \textbf{CORONA} \\
%       \hline % <-- Midrule here
%       Green box & -1.65 & -1.67 & -0.3 \\
%       Red box   & -4.8  & -3.52 & -1.13\\
%       \hline % <-- Bottomrule here
%     \end{tabular}
%   \end{center}
% \end{table}

\section{Deep learning for super-resolution}
\label{sec:deepulm}
\subsection{Ultrasound localization microscopy}
\noindent While the above described advances in front-end ultrasound processing can boost resolution, suppress clutter, and drastically improve tissue contrast, the attainable resolution of ultrasonography remains fundamentally limited by wave diffraction, i.e. the minimum distance between separable scatters is half a wavelength. Simply increasing the transmit frequency to shorten the wavelength unfortunately comes at the cost of reduced penetration depth, since higher frequencies suffer from stronger absorption compared to waves with a higher wavelength. This trade-off between resolution and penetration depth particularly hampers deep high-resolution microvascular imaging, being a cornerstone for many diagnostic applications.

Recently, this trade-off was circumvented by the introduction of \textit{ultrasound localization microscopy} (ULM) \cite{siepmann2011imaging,couture2011microbubble}. ULM leverages principles that formed the basis for the Nobel-prize-winning concept from optics of super-resolution fluoresence microscopy, and adapts these to ultrasound imaging: if individual point-sources are well-isolated from diffraction-limited scans, and their centers subsequently precisely pinpointed on a sub-diffraction grid, then the accumulation of many such localizations over time yields a super-resolved image. In optics, stochastic `blinking' of subsets of fluorophores is exploited to provide such sparse point sources. In ULM, intravascular lipid-shelled gas microbubbles fulfill this role \cite{couture2018ultrasound}. This approach permits achieving a resolution that is up to 10 times smaller than the wavelength \cite{errico2015ultrafast}. 

Since the fidelity of ULM depends on the number of localized microbubbles and the localization accuracy, it gives rise to a new trade-off that balances the required microbubble sparsity for accurate localization and acquisition time. To achieve the desired signal sparsity for straightforward isolation of the backscattered echoes, ULM is typically performed using a very diluted solution of microbubbles. On regular ultrasound systems, this constraint leads to tediously long acquisition times (on the order of hours) to cover the full vascular bed. Using an ultrafast plane-wave ultrasound system rather than regular scanning, Errico \textit{et al.} performed ultrafast ULM (uULM) in a rat brain \cite{errico2015ultrafast}. Empowered by high frame rates (500 frames per second), the acquisition time was lowered to minutes instead of hours. Ultrafast imaging indeed enables taking many snapshots of individual microubbles as they transport through the vasculature, thereby facilitating very high-fidelity reconstruction of the larger vessels. Nevertheless, mapping the full capillary bed remains dictated by the requirement of microbubbles to pass through each of the capillaries. As such, long acquisitions of tens of minutes are required, even with uULM \cite{hingot2019microvascular}. To boost the achieved coverage in a given time-span, methods that enable the use of higher concentrations can be leveraged \cite{bar2017fast,bar2018sushi,van2018super,sloun2019_deepulm,van2017sparsity}.

\subsection{Exploiting signal structure}
\label{sec:sparserecovery}
\noindent To strongly relax the constraints on microbubble concentration and therewith cover more vessels in a shorter time, standard ULM can be extended by incorporating knowledge of the measured signal structure; in particular its sparsity in a transform domain. To that end, a received contrast-enhanced image frame can be modeled as:
\begin{equation}
\mathbf{y} = \mathbf{A}\mathbf{x} +\mathbf{w},
\label{eqn:sparsemodel}
\end{equation}
where $\mathbf{x}$ is a vector which describes the sparse microbubble distribution on a high-resolution image grid, $\mathbf{y}$ is the vectorized image frame of the ultrasound sequence, $\mathbf{A}$ is the measurement matrix where each column of $\mathbf{A}$ is the point-spread-function shifted by a single pixel on the high-resolution grid, and $\mathbf{w}$ is a noise vector.

Leveraging this signal prior, i.e.,  assuming that the microbubble distribution is sparse on a sufficiently high-resolution grid (or, the number of non-zero entries in $\mathbf{x}$ is low) we can formulate the following $\ell_1$-regularized inverse problem:
\begin{equation}
\hat{\textbf{x}} = \arg\min_\mathbf{x}||\mathbf{y}-\mathbf{A}\mathbf{x}||_2^2 +\lambda||\mathbf{x}||_1,
\label{eqn:l1problem}
\end{equation}
where $\lambda$ is a regularization parameter that weighs the influence of $||\textbf{x}||_1$.

Equation \eqref{eqn:l1problem} may be solved using a numerical proximal gradient scheme such as FISTA \cite{beck2009fast}. We will discuss this FISTA-based solution in Section~IV-\ref{sec:unfolded_ulm}.
After estimating $\mathbf{x}$ for each frame, the estimates are summed across all frames to yield the final super-resolution image. 

Beyond sparsity on a frame-by-frame basis, signal structure may also be leveraged across multiple frames. To that end, a multiple-measurement vector model \cite{cotter2005sparse} and its structure in a transformed domain can be considered, e.g. by assuming that a temporal stack of frames $\mathbf{x}$ is sparse in the temporal correlation domain \cite{bar2017fast,bar2018sushi}. Considering the temporal dimension, sparse recovery may be improved by exploiting the motion of microbubbles, allowing the application of a prior on the spatial microbubble distribution through Kalman tracking \cite{solomon2018exploiting}.  

Exploiting signal structure through sparse recovery indeed enables improved localization precision and recall for high microbubble concentrations \cite{van2018super,van2017sparsity}. Unfortunately, proximal gradient schemes like FISTA typically require numerous iterations to converge (yielding a very time-consuming reconstruction process), and their effectiveness is strongly dependent on careful tuning of the optimization parameters (e.g. $\lambda$ and the step size). In addition, the linear model in \eqref{eqn:sparsemodel} is an approximation of what is actually a nonlinear relation between the microbubble distribution and the resulting beamformed and envelope-detected image frame. While this approximation is valid for microbubbles that are sufficiently far apart, the significant image-domain implications of the radio-frequency interference patterns of very closely-spaced microbubbles cannot be neglected.

\begin{figure*}[t!]
	\centering
	\includegraphics[trim=0cm 6cm 10.5cm 0cm, clip=true,width=500px]{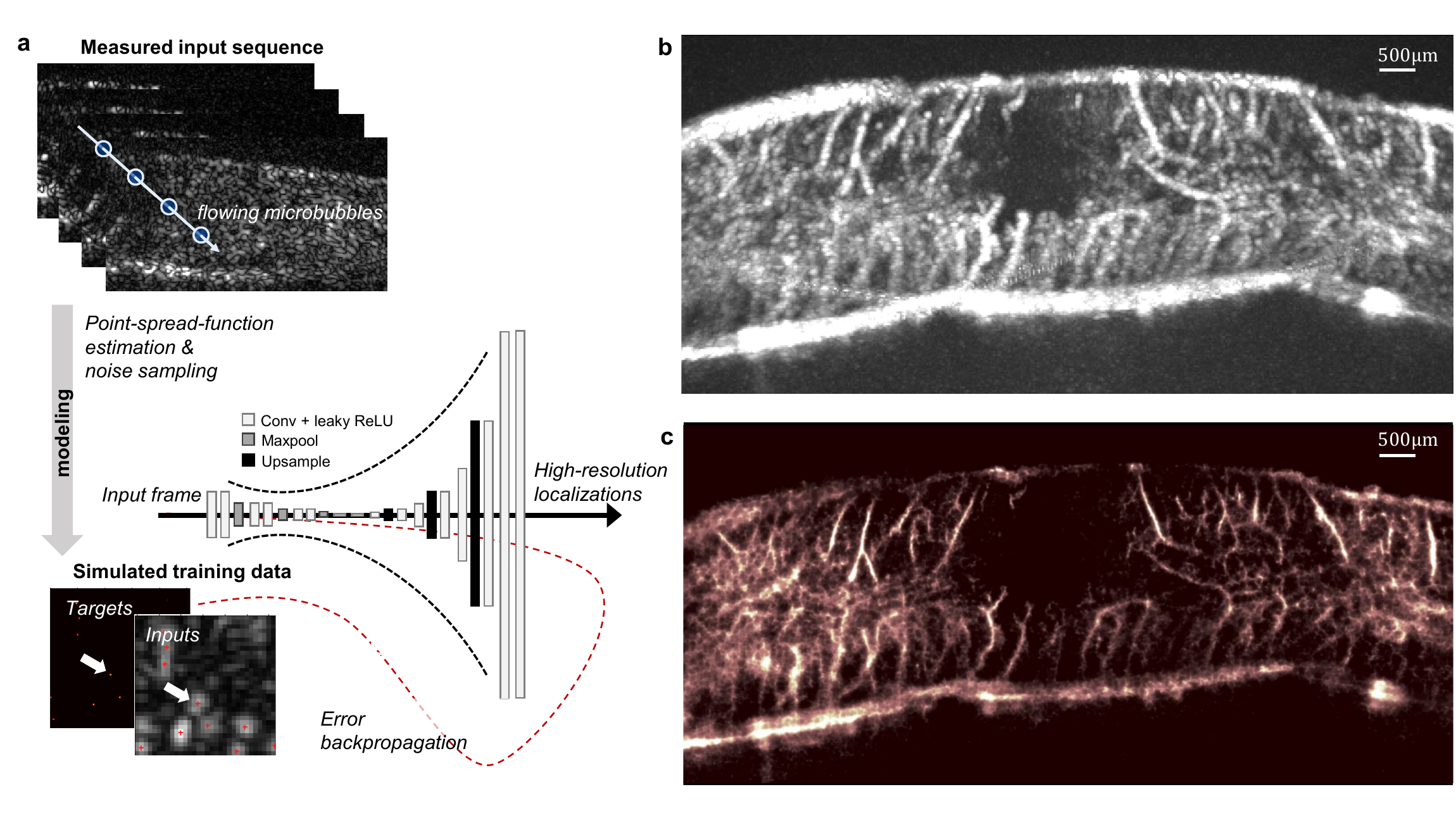}
	\caption{(a) Fast ultrasound localization microscopy through deep learning (deep-ULM) \cite{van2018super,sloun2019_deepulm}, using a convolutional neural network to map low-resolution contrast-enhanced ultrasound frames to highly resolved sparse localizations on an 8 times finer grid. The network is trained using realistic simulations of the corresponding ultrasound acquisitions, incorporating a point-spread-function estimate, the modulation frequency, pixel spacing and background noise as sampled from real datasets. (b) Standard maximum intensity projection across a sequence of frames for a rat spinal cord. (c) Corresponding deep-ULM reconstruction.}
	\label{fig:DeepULM}
\end{figure*}

\begin{figure*}[t!]
	\centering
	\includegraphics[trim=0cm 3cm 10.5cm 0cm, clip=true,width=500px]{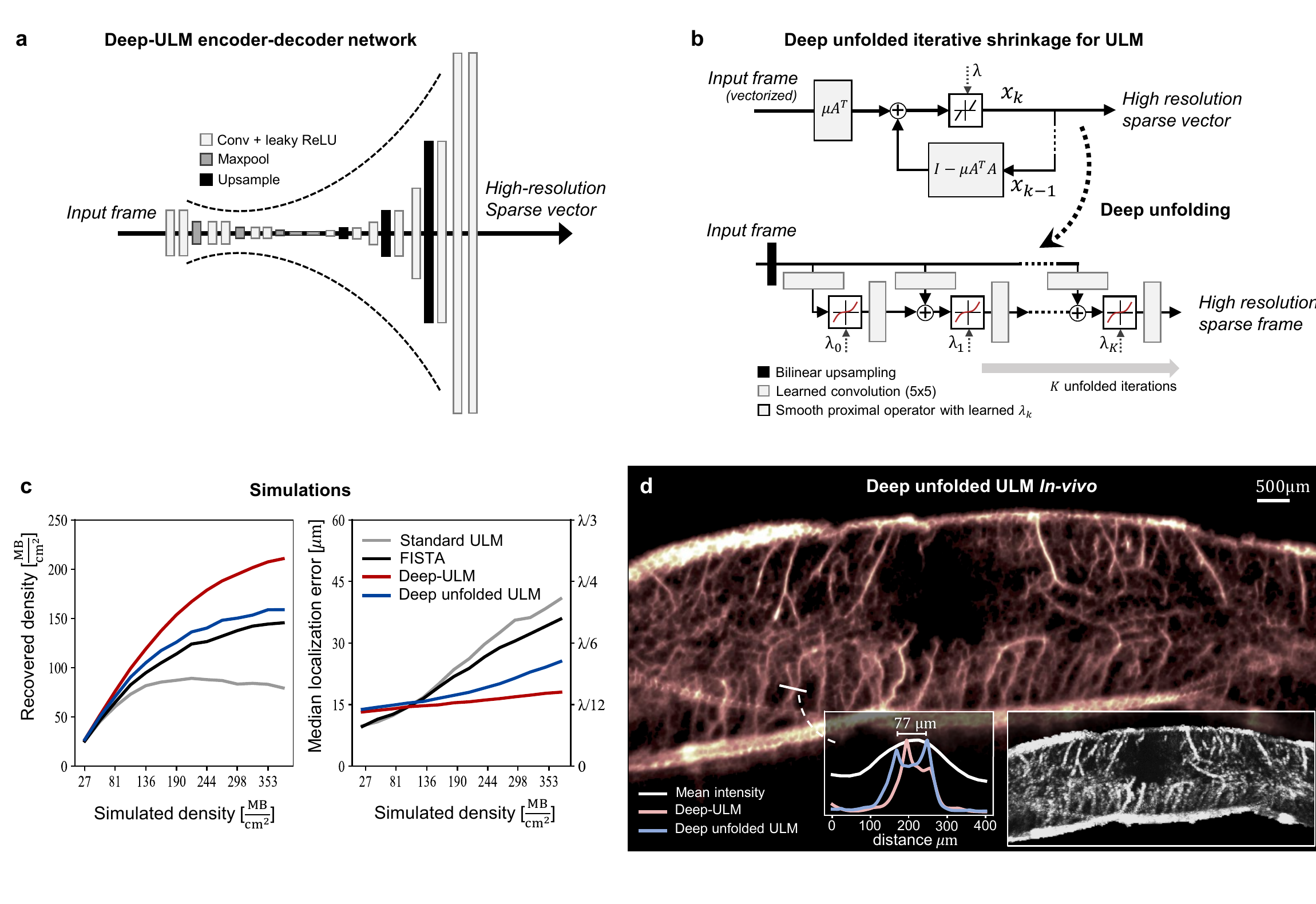}
	\caption{(a) Deep encoder-decoder architecture used in Deep-ULM \cite{van2018super,sloun2019_deepulm}, (b) Deep unfolded ULM architecture obtained by unfolding the ISTA scheme, as shown in Section~\ref{sec:deepulm}-\ref{sec:unfolded_ulm} of this paper, (c) performance comparison of standard ULM, sparse-recovery, deep-ULM and deep unfolded ULM on simulations, and (d) Deep unfolded ULM for super-resolution vascular imaging of a rat spinal cord. Both deep learning approaches outperform the other methods. While Deep-ULM shows a higher recall and slightly lower localization error as compared to deep unfolded ULM on simulation data, the latter seems to generalize better towards \textit{in-vivo} acquisitions, qualitatively yielding images with higher fidelity (see Fig.~\ref{fig:DeepULM}c for comparison).}
	\label{fig:deepISTA}
\end{figure*}

\subsection{Deep learning for fast high-fidelity sparse recovery}
\subsubsection{Encoder-decoder architectures}
In pursuit of fast and robust sparse recovery for the nonlinear measurement model, we leveraged deep learning to solve the complex inverse problem based on adequate simulations of the forward problem \cite{van2018super,sloun2019_deepulm}. This data-driven approach, named deep-ULM, harnesses a fully convolutional neural network to map a low-resolution input image containing many overlapping microbubble signals, to a high-resolution sparse output image in which the pixel intensities reflect recovered backscatter levels. This process is illustrated in Fig.~\ref{fig:DeepULM}a. The network comprises an encoder and a decoder, with the former expressing input frames in a latent representation, and the latter decoding such representation into a high-resolution output. The encoder is composed of a contracting path of 3 blocks, each block consisting of two successive $3\times3$ convolution layers and one $2\times2$ max-pooling operation. This is followed by two $3\times3$ convolutional layers and a dropout layer that randomly disables nodes with a probability of $0.5$ to mitigate overfitting. The subsequent decoder also consists of 3 blocks; the first two blocks encompassing two $5\times5$ convolution layers, of which the second has an output stride of 2, followed by $2\times2$ nearest-neighbour up-sampling. The last block consists of two convolution layers, of which the second again has an output stride of 2, preceding another 5x5 convolution which maps the feature space to a single-channel image through a linear activation function. All other activation functions in the network were leaky rectified linear units \cite{xu2015empirical}. The full deep encoder-decoder network (see  Fig.~\ref{fig:DeepULM}a) effectively scales the input image dimensions up by a factor 8, and provides a powerful model that has the capacity to learn the sparse decoding problem, while yielding simultaneous denoising through the compact latent space.

The network is trained on simulations of contrast-enhanced ultrasound acquisitions, using an estimate of the real system point-spread-function, the RF modulation frequency, and pixel spacing. Noise, clutter and artifacts were included by randomly sampling from real measurements across frames in which no microbubbles are present. Similar to \cite{nehme2018deep}, we adopt a specific loss function that acts as a surrogate for the real localization error:
\begin{equation}
    \label{eqn:loss_deepulm}
    \mathcal{L}(\mathbf{Y},\mathbf{X}_{t}|\theta)=\left\Vert f(\mathbf{Y}|\theta)-\mathbf{G}(\sigma) \ast \mathbf{X}_{t} \right\Vert_2^2+\gamma \left\Vert  f(\mathbf{Y}|\theta) \right\Vert_1,
\end{equation}
where $\mathbf{Y}$ and $\mathbf{X}_t$ are the low-resolution input and sparse super-resolution target frames, respectively, $f(\mathbf{Y}|\theta)$ is the nonlinear neural network function, and $\mathbf{G}(\sigma)$ is an isotropic Gaussian convolution kernel. Jointly, the $\ell_1$ penalty that acts on the reconstructions and the kernel $\mathbf{G}(\sigma)$ that operates on the targets, yield a loss function that increases when the reconstructed images exhibit less sparsity and when the Euclidean distances between the localizations and the targets become larger. We note that selection of the relative weighting of this sparsity penalty by $\gamma$ is less critical than the thresholding parameter $\lambda$ adopted in the sparse recovery problem \eqref{eqn:l1problem}, since the measurement model $\mathbf{A}$ (characterized by the point-spread-function) exhibits a much smaller bandwidth than $G(\sigma)$ for low values of $\sigma$ as adopted here. Consequently, the degree of bandwidth extension necessary to yield sparse outputs is less in the latter case.

Fig.~\ref{fig:DeepULM}c displays the super-resolution ultrasound reconstruction of a rat spinal cord \cite{khaing2018contrast}, qualitatively showing how deep-ULM achieves a significantly higher resolution and contrast than the diffraction-limited maximum intensity projection image (Fig. \ref{fig:DeepULM}b). Deep-ULM achieves a resolution of about 20-30~$\mu$m, being a 4-5 fold improvement compared to standard imaging with the adopted linear 15-MHz transducer \cite{van2018super}. In terms of speed, recovery on a $4096\times1328$ grid takes roughly 100 milliseconds per frame using GPU acceleration, making it about four orders of magnitude faster than a Fourier-domain implementation of sparse recovery through the FISTA proximal gradient scheme \cite{bar2018sushi}. \\

%solving the ULM inverse problem through deep learning with an $\ell1$ penalty on the outputs to enforce sparsity

\subsubsection{Deep unfolding for robust and fast sparse decoding}
\label{sec:unfolded_ulm}
While deep encoder-decoder architectures (as used in deep-ULM) serve as a general model for many regression problems and are widely used in computer vision, their large flexibility and capacity also likely make them overparameterized for the sparse decoding problem at hand. To promote robustness by exploiting knowledge of the underlying signal structure (i.e. microbubble sparsity), we propose using a dedicated and more compact network architecture that borrows inspiration from the proximal gradient methods introduced in Section~IV-\ref{sec:sparserecovery} \cite{beck2009fast}.

To do so, we first briefly describe the ISTA scheme for the  sparse decoding problem in \eqref{eqn:l1problem}:
\begin{align}
\label{eqn:ISTA}
{\mathbf{x}}^{k+1}=&\mathcal{T}_{\lambda}\left(\mathbf{x}^{k}-\mu\mathbf{A}^T\left(\mathbf{A}\mathbf{x}^{k}-\mathbf{y}\right)\right),
\end{align}
where $\mu$ determines the step size, and $\mathcal{T}_{\lambda}(\mathbf{x})_i=(|x_i|-\lambda)_+\textrm{sgn}(x_i)$ is the proximal operator of the $\ell_1$ norm. Equation \eqref{eqn:ISTA} is compactly written as:
\begin{align}
\label{eqn:ISTA2}
{\mathbf{x}}^{k+1}=&\mathcal{T}_{\lambda}\left(\mathbf{W}_1\mathbf{y} + \mathbf{W}_2\mathbf{x}^{k}\right),
\end{align}
with $\mathbf{W}_1=\mu\mathbf{A}^T$, and $\mathbf{W}_2=\mathbf{I}-\mu\mathbf{A}^T\mathbf{A}$. Similar to our approach to robust PCA in Section III-\ref{sec:RPCA}, we can \textit{unfold} this recurrent structure into a $K$-layer feedforward neural network as in LISTA (`learning ISTA') \cite{gregor2010learning}, with each layer consisting of trainable convolutions $\mathbf{W}_1^{k}$ and $\mathbf{W}_2^{k}$, along with a trainable shrinkage parameter $\lambda^{k}$. This enables learning a highly-efficient fixed-length iterative scheme for fast and robust ULM, with an optimal set of kernels and parameters per iteration, which we term \textit{deep unfolded ULM}. Different than LISTA, we avoid vanishing gradients in the `dead zone' of the proximal soft-thresholding operator $\mathcal{T}$, by replacing it by a smooth sigmoid-based soft-thresholding operation \cite{zhang2001thresholding}. An overview of this approach is given in Fig.~\ref{fig:deepISTA}b, contrasting this dedicated sparse-decoding-inspired solution with a general deep encoder-decoder network architecture in Fig.~\ref{fig:deepISTA}a. Both networks are trained on the same, synthetically generated, data. 

Tests on synthetic data show that both deep learning methods significantly outperform standard ULM and sparse decoding through FISTA for high microbubble concentrations (Fig.~\ref{fig:deepISTA}c). On such simulations, the deep encoder-decoder used in deep-ULM yields higher recall and lower localization errors compared to deep unfolded ULM. Interestingly, when applying the trained networks to \textit{in-vivo} ultrasound data, we instead observe that deep unfolded ULM yields super-resolution images with higher fidelity. Thus it is capable of translating much better towards real acquisitions than the large deep encoder-decoder network (see Figs. \ref{fig:DeepULM}c and \ref{fig:deepISTA}d for comparison). 

Our 10-layer deep unfolded ULM comprising $5\times5$ convolutional kernels has much fewer parameters (merely 506, compared to almost 700000 for the encoder-decoder scheme), therefore exhibiting a drastically lower memory footprint and reduced power consumption, in addition to achieving higher inference rates. The encoder-decoder approach requires over 4 million FLOPS for mapping a low-resolution patch of 16 by 16 pixels into a super-resolution patch of 128 by 128 pixels. The unfolded ISTA architecture is much more efficient, requiring just over 1000 FLOPS.

The lower number of trainable parameters may also explain the improved robustness and better generalization towards real data compared to its over-parameterized counterpart. On the other hand, complex image artifacts such as the strong bone reflections visible in the bottom left of Fig. \ref{fig:deepISTA}d remain more prominent using the compact unfolding scheme. 

\section{Other applications of Deep Learning in Ultrasound}
\noindent While this paper predominantly focuses on deep learning strategies for ultrasound-specific receive processing methods along the imaging chain, the initially most thriving application of deep learning in ultrasound was spurred by computer vision: automated analysis of the images obtained with traditional systems \cite{liu2019deep}. Such image analysis methods aim at dramatically accelerating (and potentially improving) current clinical diagnostics.

A classic application of ultrasonography lies in prenatal screening, where fetal growth and development is monitored to identify possible problems and aid diagnosis. These routine examinations can be complex and cumbersome, requiring years of training to swiftly identify the scan planes and structures of interest. The authors in {\cite{baumgartner2017sononet}} effectively leverage deep learning to drastically simplify this procedure, enabling real-time detection and localization of standard fetal scan planes in freehand ultrasound. Similarly, in {\cite{madani2018fast},\cite{ostvik2019real}}, deep learning was used to accelerate echocardiographic exams by automatically recognizing the relevant standard views for further analysis, even permitting automated myocardial strain imaging \cite{ostvik2018automatic}. In {\cite{song2018multi}}, a CNN was trained to perform thyroid nodule detection and recognition. Similar applications of deep learning include automated identification and segmentation of tumors in breast ultrasound \cite{chiang2018tumor}, \cite{shin2018joint}, \cite{xian2018automatic}, localization of clinically relevant B-line artifacts in lung ultrasonography \cite{sloun2019lung}, and real-time segmentation of anatomical zones on transrectal ultrasound (TRUS) scans \cite{van2019deep}. In \cite{hu2018weakly}, the authors show how such anatomical landmarks and boundaries can be exploited by a deep neural network to attain accurate voxel-level registration of TRUS and MRI.

Beyond these computer-vision applications, other learning-based techniques aim at extracting relevant medium parameters for tissue characterization. Among such approaches is data-driven elasticity imaging \cite{hoerig2017information},\cite{hoerig2018data}. In these works, the authors propose neural-network-based models that produce spatially-varying linear elastic material properties from force-displacement measurements, free from prior assumptions on the underlying constitutive models or material properties. In \cite{feigin2018deep}, a deep convolutional neural network is used for speed-of-sound estimation from (single-sided) B-mode channel data. In \cite{vishnevskiy2018image}, the authors address the problem by introducing an unfolding strategy to yield a dedicated network based on the iterative wave reflection tracking algorithm. The ability to measure speed of sound not only permits tissue characterization, but also adequate refraction-correction in beamforming.

\section{Discussion and future perspectives}
\noindent Over the past years, deep learning has revolutionized a number of domains, spurring breakthroughs in computer vision, natural language processing and beyond. In this paper we aimed to signify the potential that this powerful approach carries when leveraged in ultrasound image and signal reconstruction. We argue and show that deep learning methods profit considerably when integrating signal priors and structure, embodied by the proposed deep unfolding schemes for clutter suppression and super-resolution imaging, and the learned beamforming approaches. In addition, several ultrasound-specific considerations regarding suitable activation and loss functions were given. 

We designed and showcased a number of independent building blocks, with trained artificial agents and neural signal processors dedicated to distinct applications. Some of the presented methods operate on images (Sections~\ref{sec:RPCA} and \ref{sec:deepulm}) or IQ data (Section~\ref{sec:DopplerNet}), while others process channel data directly (Sections~\ref{sec:deep_beamforming} and \ref{sec:spectraldoppler}). A full processing chain may easily comprise a number of such components, which can be optimized holistically. This proposition enables imaging chains that are dedicated to the application and fully adaptive. %Switching from general anatomical imaging to a clutter-suppressing super-resolution contrast scheme would be merely constitute a flush of the neural weights and biases. 

Designing neural networks that can efficiently process channel data in real-time comes with a number of challenges. First, in contrast to images, channel data has a very large dynamic range and is radio-frequency modulated. This makes typical activation functions as used in image analysis (often ReLUs or hyperbolic tangents) less suited. In Section~\ref{sec:design_training}, we argue that the class of concatenated rectified linear units provides a possible alternative. Second, channel data is extremely large, in particular for large arrays or matrix transducers and when sampled at the Nyquist rate. This may be alleviated significantly by leveraging sub-Nyquist sampling schemes \cite{chernyakova2014fourier,wagner2011xampling,wagner2012compressed,mishali2011xampling,cohen2018sparse}, permitting high-end processing of low-rate channel data after (wireless) transfer to a remote (or cloud) processor. Such a new scheme, with a wireless probe that streams low-rate channel data for subsequent deep learning in the cloud, would open up many new possibilities for intelligent image formation and advanced processing in ultrasonography.

%This ability to perform beamforming from low rate samples paves the way to a wireless device where sampling can be performed in the probe due to the low rate. The compressed channel data can then be transmitted over a standard wireless channel to any computing device, or to the cloud, for further processing. This allows for transmission of channel data and paves the way to more advanced processing methods, including deep learning, which can be performed directly on the channel data rather than on the processed beamformed image.

Deep learning typically relies on vast amounts of training data. Although several approaches to make learning more data-efficient and robust have been discussed throughout this paper, a significant amount of data is still required. In the framework of supervised learning, training data typically consists of input data and desired targets. What these targets are, and how they should be obtained, depends on the application and goal. Sometimes it is for instance desirable to mimic an existing high-performance algorithm that is too complex and costly to implement in real time. Examples of this are the adaptive beamforming and spectral Doppler applications described in Sections~\ref{sec:deep_beamforming} and \ref{sec:spectraldoppler}, respectively. At other times, training data may only be obtainable through simulations or measurements on well-characterized \textit{in-vitro} phantoms. In such cases, the performance of a deep learning algorithm on \textit{in-vivo} data stands or falls with the realism of these training data and its coverage of the real-world data distribution. As shown in Section~\ref{sec:deepulm}-C2, leveraging structural signal priors in the network architecture strongly aids generalization beyond simulations.

Once trained, inference can be fast through the exploitation of high-performance GPUs. While advanced high-end imaging systems may be equipped with GPUs to facilitate the deployment of deep neural networks at the remote processor, FPGAs or ASICSs may be more appropriate for resource-limited low-power settings \cite{johansson2006ultra}. In the consumer market, small neural- and tensor-processing units (NPUs and TPUs, respectively) are enabling neural network inference at the edge \cite{jouppi2018motivation} - one can envisage a similar paradigm for front-end ultrasound processing. As such, the relevance of designing compact and efficient neural networks for memory-constrained (edge) settings is considerable and becomes particularly relevant for miniature and highly-portable ultrasound systems, where memory size, inference speed, and network bandwidth are all strictly constrained. This may be achieved by favouring (multiple) artificial agents that have very specific and well-defined tasks (Sections~\ref{sec:deep_beamforming} and \ref{sec:spectraldoppler}), as opposed to a single highly complex end-to-end deep neural network. We also showed that embedding signal priors in neural architectures permits drastically reduced memory footprints. In that context, the difference between a deep convolutional encoder-decoder network (no prior) and a deep unfolded ISTA network (structural sparsity prior) is illustrative; where the former consists of almost 700000 parameters, the latter can perform super-resolution recovery with just over 500. Additional strategies to condense large models include knowledge distillation \cite{hinton2015distilling} and parameter pruning, as well as weight quantization \cite{hubara2017quantized}.  

Once deployed in the field, artificial agents in next-generation ultrasound systems ultimately should be able to embrace the vastness of data at their disposal, to continuously learn throughout their `lifetime'. To that end, unsupervised or self-supervised learning become increasingly relevant \cite{sermanet2018time}. This holds true for many artificial intelligence applications, and extends beyond ultrasound imaging. 

The promise that deep learning holds for ultrasound imaging is significant; it may spur a paradigm shift in the design of ultrasound systems, where smart wireless probes facilitated by sub-Nyquist and neural edge computing are connected to the cloud, and with AI-driven imaging modes and algorithms that are dedicated to specific applications. Empowered by deep-learning, next-generation ultrasound imaging may become a much stronger modality with devices that continuously learn to provide better images and clinical insight, leading to improved and more widely accessible diagnostics through cost-effective, highly-portable and intelligent imaging.  

\section*{Acknowledgements}
\noindent The authors would like to thank Ben Luijten, Frederik de Bruijn and Harold Schmeitz for their contribution to the adaptive beamforming and spectral Doppler applications. They also want to thank Matthew Bruce and Zin Khaing for acquiring the spinal cord data used to evaluate the super-resolution algorithms. 

\bibliographystyle{unsrt} % Style BST file
\bibliography{Bibliography}      % Bibliography file es

\begin{IEEEbiography}[{\includegraphics[width=1in,height=1.25in,clip,keepaspectratio]{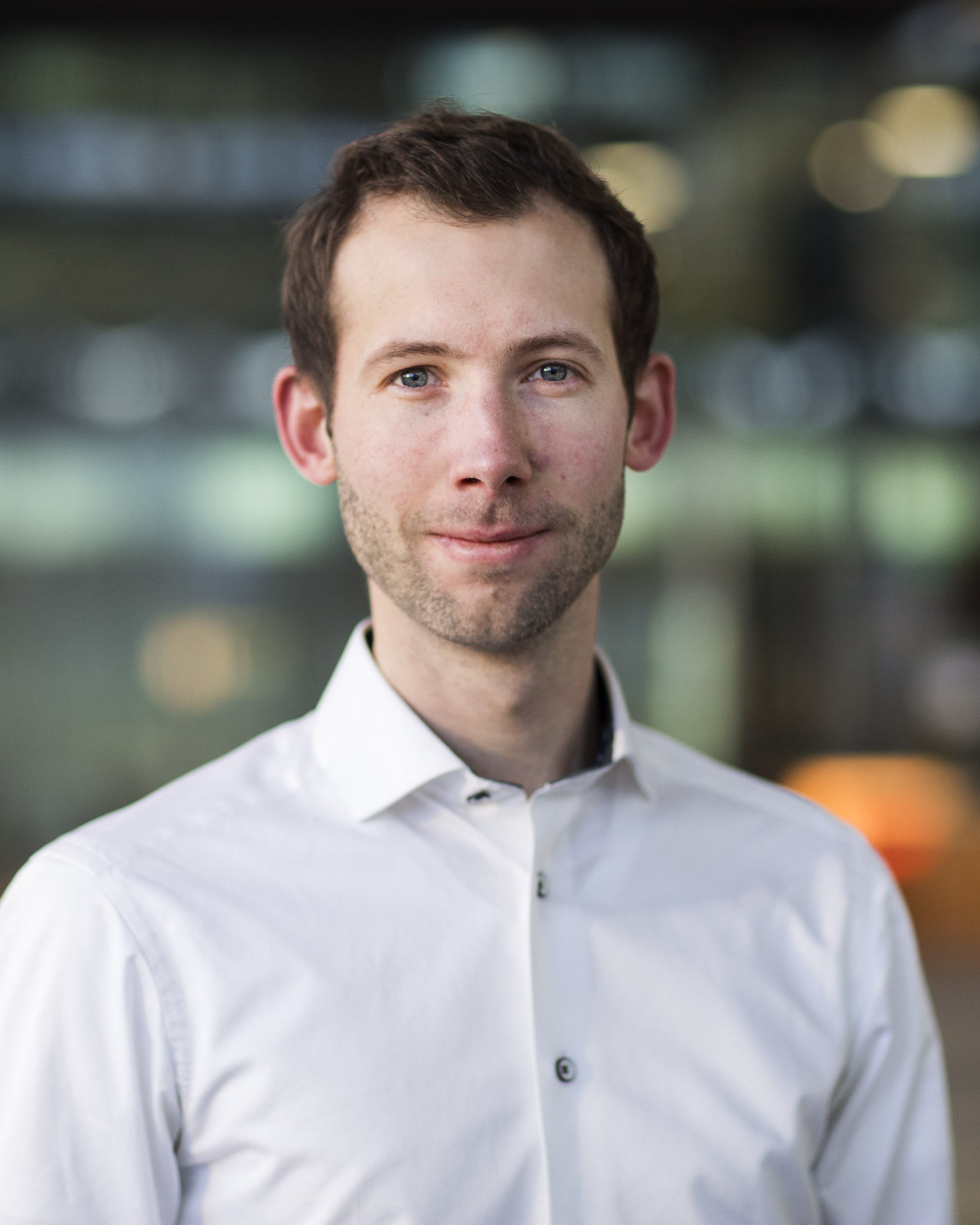}}] {Ruud J. G. van Sloun} (M'18) received the B.Sc. and M.Sc. degree (with distinction \textit{cum laude}) in Electrical Engineering from the Eindhoven University of Technology, The Netherlands, in 2012 and 2014, respectively. In 2018, he received the Ph.D. degree (with distinction \textit{cum laude}) from the Eindhoven University of Technology. Since then, he has been an Assistant Professor at the department of Electrical Engineering, Eindhoven University of Technology, The Netherlands. He is also a Visiting Professor at the department of Mathematics and Computer Science, Weizmann institute of Science, Israel. His research interests include artificial intelligence and deep learning for front-end signal processing, model-aware deep learning, compressed sensing, ultrasound imaging, and signal \& image analysis.
\end{IEEEbiography}

\begin{IEEEbiography}[{\includegraphics[width=1in,height=1.25in,clip,keepaspectratio]{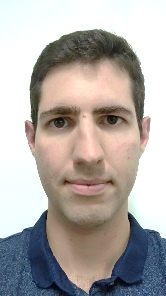}}] {Regev Cohen} (GS'16) received the B.Sc. degree (\textit{summa cum laude}) in electrical engineering from the Technion-Israel Institute of Technology, Haifa, Israel, in 2015, where he is currently pursuing the Ph.D. degree. His current research interests include theoretical aspects of signal processing, sampling theory, compressed sensing, optimization methods, sparse array design, deep learning, and advanced signal processing methods for ultrasonic imaging. Mr. Cohen received the Meyer Foundation Excellence Award and Elias-Perlmutter Award in 2015. In 2017, he was awarded with the Israel and Debora Cederbaum Scholarship.
\end{IEEEbiography}

\begin{IEEEbiography}[{\includegraphics[width=1in,height=1.25in,clip,keepaspectratio]{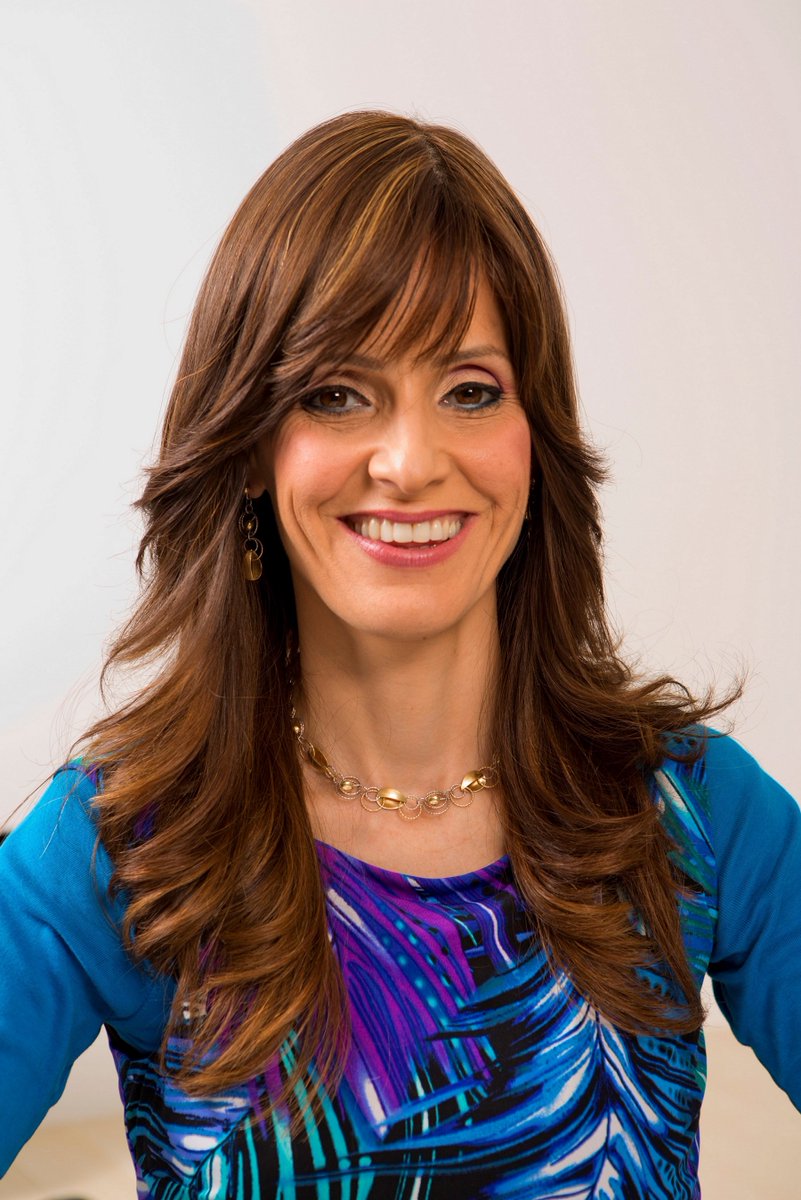}}] {Yonina C. Eldar} (S'98-M'02-SM'07-F'12) received the B.Sc. degree in Physics in 1995 and the B.Sc. degree in Electrical Engineering in 1996 both from Tel-Aviv University (TAU), Tel-Aviv, Israel, and the Ph.D. degree in Electrical Engineering and Computer Science in 2002 from the Massachusetts Institute of Technology (MIT), Cambridge.

She is currently a Professor in the Department of Mathematics and Computer Science, Weizmann Institute of Science, Rechovot, Israel. She was previously a Professor in the Department of Electrical Engineering at the Technion, where she held the Edwards Chair in Engineering. She is also a Visiting Professor at MIT, a Visiting Scientist at the Broad Institute, and an Adjunct Professor at Duke University and was a Visiting Professor at Stanford. She is a member of the Israel Academy of Sciences and Humanities (elected 2017), an IEEE Fellow and a EURASIP Fellow. Her research interests are in the broad areas of statistical signal processing, sampling theory and compressed sensing,  learning and optimization methods, and their applications to medical imaging, biology and optics.

Dr. Eldar has received many awards for excellence in research and teaching, including the  IEEE Signal Processing Society Technical Achievement Award (2013), the IEEE/AESS Fred Nathanson Memorial Radar Award (2014), and the IEEE Kiyo Tomiyasu Award (2016).
She was a Horev Fellow of the Leaders in Science and Technology program at the Technion and an Alon Fellow. She received the Michael Bruno Memorial Award from the Rothschild Foundation, the Weizmann Prize for Exact Sciences, the Wolf Foundation Krill Prize for Excellence in Scientific Research, the Henry Taub Prize for Excellence in Research (twice), the Hershel Rich Innovation Award (three times), the Award for Women with Distinguished Contributions, 
the Andre and Bella Meyer Lectureship, the Career Development Chair at the Technion, the Muriel \& David Jacknow Award for Excellence in Teaching, and the Technion's Award for Excellence in Teaching (two times).  She received several best paper awards and best demo awards together with her research students and colleagues including the SIAM outstanding Paper Prize, the UFFC Outstanding Paper Award, the Signal Processing Society Best Paper Award and the IET Circuits, Devices and Systems Premium Award, and was selected as one of the 50 most influential women in Israel.

She was a member of the Young Israel Academy of Science and Humanities and the Israel Committee for Higher Education. She is the Editor in Chief of Foundations and Trends in Signal Processing, a member of the IEEE Sensor Array and Multichannel Technical Committee and serves on several other IEEE committees. In the past, she was a Signal Processing Society Distinguished Lecturer, member of the IEEE Signal Processing Theory and Methods and Bio Imaging Signal Processing technical committees, and served as an associate editor for the IEEE Transactions On Signal Processing, the EURASIP Journal of Signal Processing, the SIAM Journal on Matrix Analysis and Applications, and the SIAM Journal on Imaging Sciences. She was Co-Chair and Technical Co-Chair of several international conferences and workshops.

She is author of the book ``Sampling Theory: Beyond Bandlimited Systems'' and co-author of the books ``Compressed Sensing'' and ``Convex Optimization Methods in Signal Processing and Communications'', all published by Cambridge University Press.

\end{IEEEbiography}

\end{document}